\documentclass[pra,twocolumn,showpacs,preprintnumbers,amsmath,amssymb,aps,superscriptaddress]{revtex4-2}
\usepackage{color}
\usepackage{bm, graphicx, amsmath}
\usepackage{hyperref}
\usepackage{comment}

\usepackage{braket}

\begin{document}
\title{Broadcasting single-qubit and multi-qubit-entangled states: \\
authentication, cryptography, and distributed quantum computation}

\author{Hiroki Sukeno}
\affiliation{C. N. Yang Institute for Theoretical Physics and Department of Physics and Astronomy, State University of New York at Stony Brook, NY 11794-3840, USA}
\author{Tzu-Chieh Wei}
\affiliation{C. N. Yang Institute for Theoretical Physics and Department of Physics and Astronomy, State University of New York at Stony Brook, NY 11794-3840, USA}
\author{Mark Hillery}
\affiliation{Department of Physics and Astronomy, Hunter College of the City University of New York, 695 Park Avenue, New York, NY 10065, USA}
\affiliation{Graduate Center of the City University of New York, 365 Fifth Avenue, New York, NY 10016, USA}
\author{J\'{a}nos A. Bergou}
\affiliation{Department of Physics and Astronomy, Hunter College of the City University of New York, 695 Park Avenue, New York, NY 10065, USA}
\affiliation{Graduate Center of the City University of New York, 365 Fifth Avenue, New York, NY 10016, USA}
\author{Dov Fields}
\affiliation{DEVCOM Army Research Laboratory, Adelphi, MD 20783, USA}
\author{Vladimir S. Malinovsky}
\affiliation{DEVCOM Army Research Laboratory, Adelphi, MD 20783, USA}

\begin{abstract}
Quantum entanglement assisted with measurements provides various pathways to communicate information to parties within a network. In this work, we  generalize a previous broadcasting protocol and present schemes to broadcast product and multi-partite entangled  
quantum states, where in the latter case  the sender can remotely add phase gates or abort distributing the states.  We first focus on the broadcasting of product  quantum states in a network, and generalize the basic protocol to include an arbitrary basis rotation and allow for  multiple receivers and senders.  We show how to add and delete senders from the network. The generalization also includes the case where  a phase to be applied to the broadcast states is not known in advance but is provided to a sender encoded in another quantum state. Applications of broadcasting product states include  authentication and three-state quantum cryptography.
In the second part, we study the distribution of a single multi-qubit state shared among several receivers entangled with multi-qubit phase gates, which includes the graph states as an example.
We show that by coordinating with the sender, the receivers can assist in performing remote, distributed measurement-based quantum computation with the Pauli $X$ basis measurement alone.
As another application of this, we discuss the distribution of the multi-qubit Greenberger-Horne-Zeilinger state.
\end{abstract}

\maketitle

\section{Introduction}
In a quantum network, quantum information, encoded in the form of qubits or perhaps qudits, can be sent between users either by sending the qubits directly or making use of shared entanglement (such as in  the teleportation protocol)~\cite{kimble}.  If the same information is to be sent to a number of different users, the No-Cloning Theorem presents a constraint on one's ability to do so for general quantum states.  Despite this, several schemes for broadcasting quantum information have been investigated, but they typically require one entanglement bit (ebit) per qubit for each receiving party~\cite{Yu1,Yu2,Zhou,Zhao}.

An alternative, if one has knowledge of the state to be distributed, is a version of remote state preparation~\cite{Lo,Bennett,Leung,Hayden,Zukowski}.  For sending states to a single receiver, remote state preparation has a smaller classical communication cost than teleportation.  If one goes a step further and restricts the set of states to be sent, then it can be adapted for broadcasting quantum information~\cite{pati,hillery}.  The protocol in~\cite{pati} made use of dark states, while that in~\cite{hillery} used entangled Dicke states.

In Ref.~\cite{hillery} we showed how qubit states of the form $e^{i\theta}\alpha |0\rangle + e^{-i\theta}\beta |1\rangle$, where $\alpha$ and $\beta$ are fixed but $\theta$ can be varied by the sender, can be simultaneously sent to a number of different receivers.  Let us briefly summarize how this works for two receivers, and refer to this as the basic broadcast protocol (BBP). Alice prepares a state consisting of one qutrit and two qubits in the form,
\begin{eqnarray}
\label{template}
|\Psi^{(1,2)}\rangle_{abc} & = & \alpha^{2}|0\rangle_{a} |00\rangle_{bc} + \beta^{2} |1\rangle_{a} |11\rangle_{bc} \nonumber \\
& & + \alpha\beta |2\rangle_{a} (|01\rangle_{bc} + |10\rangle_{bc} ),
\end{eqnarray}
and transmits the two qubits to Bob and Charlie, separately. One can regard this state as the starting point shared among Alice, Bob and Charlie.
Note that superscript $(1,2)$ denotes that we have one sender and two receivers.
Later we generalize this to $|\Psi^{(M,N)}\rangle$ when we consider the case with  $M$ senders and $N$ receivers.

Alice then applies the operator $U_{a}(\theta)$ to her qutrit, where $U_{a}|0\rangle_{a} = e^{2i\theta}|0\rangle_{a}$, $U_{a}|1\rangle_{a} = e^{-2i\theta}|1\rangle_{a}$, and $U_{a}|2\rangle_{a} = |2\rangle_{a}$.  She then measures her qutrit in the basis
\begin{eqnarray}
\label{basis}
|u_{0}\rangle_{a} & = & \frac{1}{\sqrt{3}}( |0\rangle_{a} + |1\rangle_{a} + |2\rangle_{a}) \nonumber \\
|u_{1}\rangle_{a} & = & \frac{1}{\sqrt{3}}( e^{2\pi i/3}|0\rangle_{a} + e^{-2\pi i/3}|1\rangle_{a} + |2\rangle_{a}) \nonumber \\
|u_{2}\rangle_{a} & = & \frac{1}{\sqrt{3}}( e^{-2\pi i/3}|0\rangle_{a} + e^{2\pi i/3}|1\rangle_{a} + |2\rangle_{a}) ,
\end{eqnarray}
and sends the result of her measurement $m_a$ (which can be 0, 1 or 2) to Bob and Charlie.  Each of them applies a correction operator, which depends on $m_a$, to their respective qubit.  Specifically, for $m_a=0$, there is no correction needed. For $m_a=1$, Bob and Charlie each applies $U_C$, with $U_{C}|0\rangle_{a} = e^{-i \pi/3}|0\rangle_{a}$, $U_{C}|1\rangle_{a} = e^{-i\pi/3}|1\rangle_{a}$; for $m_a=2$, they apply $U_C^{-1}$ instead.  The resulting state of Bob and Charlie is the same qubit state at stated in the beginning, i.e., $e^{i\theta}\alpha |0\rangle + e^{-i\theta}\beta |1\rangle$.

The procedure is both a restriction and an extension of remote state preparation.  It is a restriction, because the set of transmitted states is not the entire qubit space, since $\alpha$ and $\beta$ are fixed.  It is an extension because the states are sent to multiple receivers. In Ref.~\cite{hillery},  we also showed how a value of $\theta$ unknown to the sender, but encoded in a quantum state, can be sent probabilistically to multiple receivers.  Here, we would like to extend those results and generalize our BBP.  First, we will show how Alice can apply a more general set of operations to the remote qubits.  Rather than $e^{i\theta}|0\rangle\langle 0| + e^{-i\theta} |1\rangle\langle 1|$ she can apply $e^{i\theta}|s_{0}\rangle\langle s_{0}| + e^{-i\theta} |s_{1}\rangle\langle s_{1}|$, where $\langle s_{0}|s_{1}\rangle = 0$.  Next, we will show how to incorporate multiple senders into the procedure, and how senders can be added or deleted from the network.  We will then look further into the situation when Alice does not know $\theta$, which is encoded into a quantum state she receives.  There are both probabilistic and approximate procedures that allow her  to transfer the general unknown $\theta$ into the qubits held by Bob and Charlie.  Additionally, we will provide two more applications of these protocols: the first application  enables the generation of sequences that can be used for the purpose of authentication, and the second  is a three-state quantum key distribution (QKD) procedure.

Furthermore, as an extension of our BBP,
we also consider distributing a single multi-qubit
state shared among several receivers, including the previous product broadcast qubit states extended by additional multi-qubit entangling phase gates and general stabilizer states (such as graph states~\cite{Hein})).
We explain how some of these multi-partite entangled states can be distributed in two different ways. 
It turns out that in both methods, the sender can teleport phases chosen after the distribution to receivers.
The latter method especially enables a remote, distributed measurement-based quantum computation (MBQC)~\cite{MBQC-PRA,MBQC-PRL}, with the sender performing axis adaptation and receivers performing only $X$-basis measurement. 
Furthermore, we describe how the GHZ state \cite{GHZ} can be distributed.

The structure of the remaining paper is as follows. 
In Sec.~\ref{sec:generalization-of-broadcasting}, we extend the protocol and show that Alice can apply more general operations to the remote qubits. We then explain the generalization to the protocol with multiple senders.
In Sec.~\ref{sec:applications}, we discuss the application of protocol where Alice sends unknown phases to receivers.
We then discuss applications to the authentication and the three-state QKD.
In Sec.~\ref{sec:broadcasting-a-distributed-stabilizer-state}, we explain our method to broadcast a distributed stabilizer state.
In Sec.~\ref{sec:applications-of-distributed-stabilizer-states}, we discuss applications of distributed stabilizer states.
Sec.~\ref{sec:conclusion} is devoted to conclusions.

\section{Extensions of broadcasting scheme for single qubit states}
\label{sec:generalization-of-broadcasting}

In this section, we present a couple of generalizations of our BBP. First, we extend it to the case where the basis in the broadcast state can be arbitrary. Then, we generalize the BBP to multiple senders and multiple receivers.

\subsection{Extension of BBP with general single-qubit unitaries}

Consider the following scenario.  Alice possesses two qubits, both of which are in the state $|\psi\rangle$, which she may or may not know.  She wants to transmit qubits to Bob and Charlie in the rotated basis, $(e^{i\theta} |s_{0}\rangle\langle s_{0}| + e^{-i\theta} |s_{1}\rangle\langle s_{1}|)|\psi\rangle$, where she can vary the phase angle $\theta$ at a later time.  Here, $\{ |s_{0}\rangle , |s_{1}\rangle\}$ is an orthonormal qubit basis.  Let $T$ be the unitary operator that maps the $\{ |0\rangle , |1\rangle \}$ basis to the $\{ |s_{0}\rangle , |s_{1}\rangle\}$ basis, that is $T|j\rangle = |s_{j}\rangle$ for $j=0,1$, and suppose that $|\psi\rangle = \mu |s_{0}\rangle + \nu |s_{1}\rangle$.  Alice first applies $T^{-1}$ to each of her qubits and subjects them to the circuit in Fig.\ 1.  The top line is a qutrit, and the bottom two lines are the two qubits.  The gates labelled $V$ are Controlled-Shift gates, where $V$ is the shift, in which the control is a qubit and the target a qutrit. If the control is in the state $|0\rangle$, nothing happens to the target, and if the control is in the state $|1\rangle$, then $V$ is applied to the qutrit, where $V|j\rangle = |(j-1){\rm mod}\, 3\rangle$ for $j=0,1,2$, and the subtraction is modulo $3$. She then applies $T\otimes T$ to the qubits at the output of the circuit.  The resulting state is
\begin{eqnarray}
\label{template2}
 \mu^{2}|0\rangle_{a} |s_{0},s_{0}\rangle_{bc} + \nu^{2} |1\rangle_{a} |s_{1},s_{1}\rangle_{bc} \nonumber \\
 + \mu\nu |2\rangle_{a} (|s_{0},s_{1}\rangle_{bc} + |s_{1},s_{0}\rangle_{bc} ) .
\end{eqnarray}

The qubits are then sent to Bob and Charlie, and at a later time $U_{a}(\theta)$ can be applied to the qutrit as in the original protocol. After that, the qutrit is then measured and Bob and Charlie perform any necessary correction procedures according to Alice's measurement outcome.

\begin{figure}
\begin{center}
\includegraphics[width=24em]{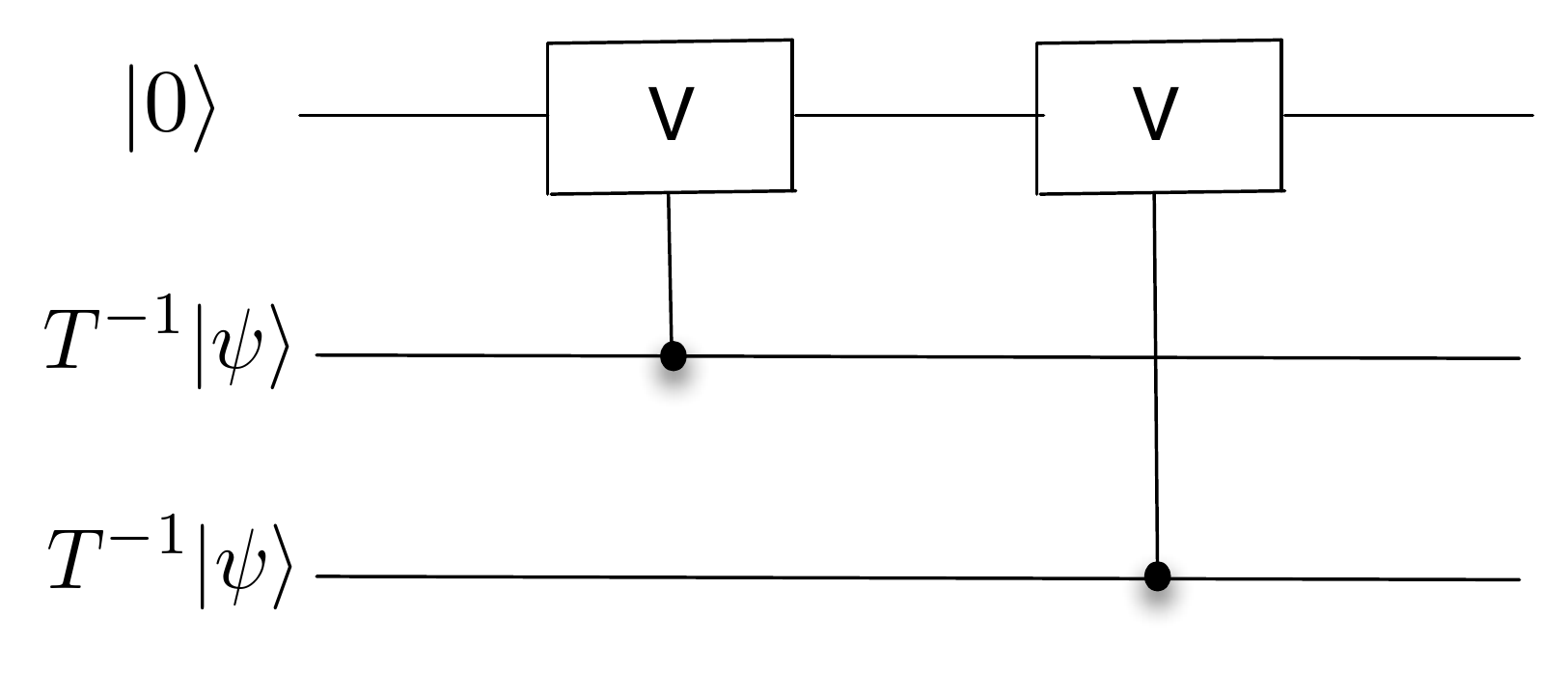}
\end{center}
\caption{Circuit for producing the state in  Eq.\ (\ref{template2}).  The top line is a qutrit, initially in the state $|0\rangle$, and the bottom two lines are qubits.  The operation $T$ is applied to the qubits at the output.}
\end{figure}

\subsection{Extension to multiple senders}
In this part, we are generalizing our BBP to the case of multiple senders and multiple receivers.

\subsubsection{Example: two senders and two receivers}
\label{sec:twosenders}

To begin with, we now study the example of two senders and two receivers.  We have two Alices, Alice 1 and Alice 2, who are the senders, and Bob and Charlie who are the receivers.  Alice 1 and Alice 2 have qutrits, and Bob and Charlie have qubits.  We start with the entangled two-qutrit and two-qubit state 
\begin{eqnarray}
\label{multsend}
|\Psi^{(2,2)}\rangle & = & \alpha^{2} |0\rangle_{a1}\otimes |0\rangle_{a2}\otimes|00\rangle_{bc}  \nonumber \\
& & + \alpha\beta  |1\rangle_{a1} \otimes |1\rangle_{a2} \otimes(|01\rangle_{bc} + |10\rangle_{bc})  \nonumber \\
& & +\beta^{2} |2\rangle_{a1} \otimes |2\rangle_{a2} \otimes |11\rangle_{bc},
\end{eqnarray}
which can be prepared locally, e.g., by Alice 1, and then distribute the corresponding qutrit and qubits to others.
Alice 1 applies the operator $U_{1}(\theta_{1})$ to her qutrit, where
\begin{eqnarray}
&U_{1}(\theta_{1})|0\rangle_{a1}=e^{i\theta_{1}}|0\rangle_{a1}, & \hspace{5mm}  U_{1}(\theta_{1})|1\rangle_{a1} = |1\rangle_{a1},  \nonumber \\
&U_{1}(\theta_{1})|2\rangle_{a1}= e^{-i\theta_{1}}|2\rangle_{a1}.   &
\end{eqnarray}
Alice 2 applies $U_{2}(\theta_{2})$ to her qutrit, and it has the same action as $U_{1}$ except with $\theta_{1}$ replaced by $\theta_{2}$.  We find that
\begin{eqnarray}
&&\big(I_{bc}\otimes U_{1}(\theta_{1})\otimes U_{2}(\theta_{2})\big)|\Psi^{(2,2)}\rangle \hspace{2cm} \nonumber \\
&& =  \alpha^{2} e^{i(\theta_{1}+\theta_{2})}|0\rangle_{a1}\otimes |0\rangle_{a2}\otimes |00\rangle_{bc} \nonumber \\
&& \quad+ \alpha\beta  |1\rangle_{a1} \otimes |1\rangle_{a2} \otimes(|01\rangle_{bc} + |10\rangle_{bc})  \nonumber \\
&& \quad + \beta^{2} e^{-i(\theta_{1}+\theta_{2})}  |2\rangle_{a1} \otimes |2\rangle_{a2} \otimes |11\rangle_{bc}.
\end{eqnarray}
Both Alice 1 and Alice 2 now measure their qutrits in the $\{ |u_{j}\rangle\, |\, j=0,1,2\}$ basis, where
\begin{equation}
|u_{j}\rangle = \frac{1}{\sqrt{3}}\sum_{l=0}^{2} e^{2\pi i lj/3} |l\rangle  .
\end{equation}
If Alice 1 gets $j$ and Alice 2 gets $k$, then the resulting joint state of Bob and Charlie is
\begin{align} 
&|\Psi^{(2,2)}\rangle  \xrightarrow[\text{measure}]{U_1(\theta_1), U_2(\theta_2)} |\Phi_{(j,k)}\rangle, 
\end{align}
where
\begin{eqnarray}
&& |\Phi_{(j,k)}\rangle  = \left[ \alpha e^{i(\theta_{1}+\theta_{2})/2} |0\rangle_{b} + \beta e^{-2\pi i (j+k)/3} e^{-i(\theta_{1}+\theta_{2})/2} |1\rangle_{b} \right]   \nonumber \\ 
&&\quad\otimes \left[ \alpha e^{i(\theta_{1}+\theta_{2})/2} |0\rangle_{c} 
+ \beta e^{-2\pi i (j+k)/3} e^{-i(\theta_{1}+\theta_{2})/2} |1\rangle_{c} \right].
\end{eqnarray}
If Alice 1 and Alice 2 tell Bob and Charlie what their measurement results were, then Bob and Charlie can each apply a unitary operator to their qubits to remove the phase factor $e^{-2\pi i (j+k)/3}$.  Thus, Bob and Charlie each receive a qubit that contains contributions from both Alice 1 and Alice 2.  The procedure can be adapted so that only one Alice determines the sent qubits.  For example, if Alice 2 does not apply $U_{2}(\theta_{2})$, then $\theta_{2}=0$, and the sent qubits are determined only by Alice 1.  However, to send these qubits to Bob and Charlie, the cooperation of Alice 2 is required; she has to measure her qutrit and send the result to Bob and Charlie.

\subsubsection{General case:  $M$ senders and $N$ receivers}
This procedure can be extended to the case of $M$ senders and  $N$ receivers.  We start with a state consisting of $M$ qudits with $N+1$ levels and $N$ qubits 
\begin{equation}
|\Psi^{(M,N)}\rangle = \sum_{k=0}^{N} \alpha^{k} \beta^{N-k} \left( \begin{array} {c} N \\ k \end{array}\right)^{1/2}
\left(\prod_{j=1}^{M}  |k\rangle_{aj} \right)|k;N-k\rangle .
\end{equation}
Here $|k\rangle_{aj}$ is the state of the qudit of Alice $j$, and $|k;N-k\rangle$ is an $N$-qubit state, which is a normalized completely symmetric state in which $k$ of the qubits are in the state $|0\rangle$ and $N-k$ are in the state $|1\rangle$,
\begin{equation}
 |k;N-k\rangle \equiv\left( \begin{array} {c} N \\ k \end{array}\right)^{-1/2}{\rm Symmetrize} \big\{\underbrace{|0\cdots0}_{k}\underbrace{1\cdots1\rangle}_{N-k}\big\}.
\end{equation}
These $N$-qubit states are also known as Dicke states.  Each Alice $j$ has one of the qudits, and each of the receivers has one of the qubits.  Alice $j$ can now choose an angle, $\theta_{j}$, by applying the operator $U_{aj}$ to her qudit, where
\begin{equation}
U_{aj}(\theta_{j} )|k\rangle_{aj} = e^{i(2k-N)\theta_{j} }|k\rangle_{aj} .
\end{equation}
The resulting state is
\begin{eqnarray}
\prod_{j=1}^{M}U_{aj} |\Psi^{(M,N)}\rangle & = & \sum_{k=0}^{N} \Big(\alpha\prod_{j^{\prime}=1}^{M}e^{i\theta_{j^{\prime}}}\Big)^{k}  \Big(\beta \prod_{j^{\prime\prime}=1}^{M} e^{-\theta_{j^{\prime\prime}}}\Big)^{N-k}  \nonumber\\
&&\times
\left( \begin{array} {c} N \\ k \end{array}\right)^{1/2}\prod_{j_3=1}^{M}|k\rangle_{aj_3} |k;N-k\rangle.
\end{eqnarray}
Each Alice $j$ now  measures her qudit in the basis
\begin{equation}
|u_{n}\rangle = \frac{1}{\sqrt{N+1}} \sum_{k=0}^{N} e^{2\pi ink/(N+1)} |k\rangle ,
\end{equation}
where $n=0,1, \ldots N$.  If Alice $j$ obtains the result $|u_{n_{j}}\rangle$ the unnormalized $N$ qubit state is
\begin{align}
&|\Psi^{(M,N)}\rangle 
\xrightarrow[\text{measure}]{U_j(\theta_j)^{\otimes M}} 
|\Phi_{\bar{n}}\rangle   \ ,  \nonumber \\
&|\Phi_{\bar{n}}\rangle 
=  \prod_{l=1}^{N}\left( \alpha e^{i\sum_{j=1}^{M} \theta_{j}} e^{-2\pi i (\sum_{j=1}^{M} n_{j})/(N+1)} |0\rangle_{l} \right. \nonumber \\
\label{eq:general-distributed-state}
& \qquad \qquad  \left. + \beta e^{-i\sum_{j=1}^{M} \theta_{j}} |1\rangle_{l} \right)  ,
\end{align}
where $\bar{n}=(n_{1},n_{2},\ldots n_{M})$.  Each Alice then broadcasts the result of her measurement, and each of the parties applies the correction operator, $U_{\bar{n}}$ to their qubit, where $U_{\bar{n}}|0\rangle = \exp [2\pi i (\sum_{j=1}^{M} n_{j})/(N+1) ] |0\rangle$ and $U_{\bar{n}}|1\rangle = |1\rangle$.  This will result in each party possessing the state $\alpha e^{i\sum_{j=1}^{M} \theta_{j}} |0\rangle + \beta e^{-i\sum_{j=1}^{M} \theta_{j}} |1\rangle$.

If only one sender is to send a message, that sender performs the corresponding unitary operation, and the others do nothing.  However, all senders do have to measure their qudits and broadcast their results in order to complete the procedure.

\subsubsection{Adding senders to the network: $M \rightarrow M+1$}
It is possible to add senders if one of the original senders shares a fully entangled state with them.  
The purpose of this subsection is to present a procedure to generate $|\Psi^{(M+1,N)}\rangle$ from $|\Psi^{(M,N)}\rangle$ ($M \geq 1$).
Suppose we start with the state
\begin{equation}
|\Psi^{(M,N)}\rangle \otimes  \frac{1}{\sqrt{N+1}} \sum_{\ell=0}^{N} |\ell\rangle_{aM^{\prime}} |\ell \rangle_{a(M+1)} .
\end{equation}
Qudits $aM$ 
and $aM^{\prime}$ are both held by Alice $M$, and $a(M+1)$ is held by the person we want to add as the additional sender, Alice $M+1$.  Alice $M$  sends them through a Controlled-Shift gate (written as $CX_{c,t}$), with $aM$ as the control and $aM^{\prime}$ as the target.  The Controlled-shift gate acts as $|k\rangle_{aM}|\ell\rangle_{aM^{\prime}} \rightarrow |k\rangle_{aM}|\ell+k\rangle_{aM^{\prime}}$, where the  addition is modulo $N+1$.  Alice $M$  now measures $aM^{\prime}$ in the computational basis.  If she gets $|j\rangle_{aM^{\prime}}$,  
\begin{align}
&CX_{aM,aM'} \Big( 
|\Psi^{(M,N)}\rangle \otimes  \frac{1}{\sqrt{N+1}} \sum_{\ell=0}^{N} |\ell\rangle_{aM^{\prime}} |\ell \rangle_{a(M+1)} \Bigr) \nonumber \\
& \xrightarrow{\text{measure}}
\sum_{k=0}^{N} \alpha^{k} \beta^{N-k} \left( \begin{array} {c} N \\ k \end{array}\right)^{1/2} \nonumber \\ 
& \qquad \qquad \times \prod_{j=1}^{M}  |k\rangle_{aj} \otimes |j-k \rangle_{a(M+1)} \otimes
|k;N-k\rangle.  
\end{align}
Alice $M$ sends the result of her measurement to Alice $M+1$, who can now apply a local unitary operator to map $|j-k\rangle_{a(M+1)} \rightarrow |k\rangle_{a(M+1)}$,
then the resulting state is $|\Psi^{(M+1,N)}\rangle$. 

A sender can delete herself by simply measuring her qutrit, and sending the result of the measurement to Bob$_1$, Bob$_2$, ... Bob$_N$, who can then each apply a unitary operator to remove the phase factor resulting from the measurement by the deleting party.

\section{Further generalization and applications of broadcasting single-qubit states}
\label{sec:applications}
Here, we give several applications of broadcasting single-qubit states. The first is an extension to send an unknown phase encoded in a quantum state. The second is an application of our broadcasting protocol to authentication. The last is an application to quantum key distribution using three states.
\subsection{Sending unknown phases}

Let us return to the case of one sender, Alice, who wants to send Bob and Charlie a state with an unknown phase.  This was studied in \cite{hillery}.  There Alice had one qutrit, which is entangled with the qubits of Bob and Charlie.  The phase was encoded in the qutrit state 
\begin{equation}
|\Phi\rangle_{d} = \frac{1}{\sqrt{3}} (e^{2i\theta} |0\rangle_{d} + |1\rangle_{d} + e^{-2i\theta} |2\rangle_{d})  .
\end{equation}
A combination of a unitary operation and measurement allowed Alice to send the state $e^{i\theta}\alpha |0\rangle + e^{-i\theta} \beta |1\rangle$, where, as usual, $\alpha$ and $\beta$ are fixed, to Bob and Charlie with a probability of $1/3$.  Here we would like to expand on that result.  First, we will
show that if $\theta$ is in the set $\{ \pi k/K\, | \, k=0,1,\ldots K-1\}$ and encoded in a $K$-level system, where $K$ is fixed, we can  send the state $e^{i\theta}\alpha |0\rangle + e^{-i\theta} \beta |1\rangle$ to Bob and Charlie with certainty at the expense of using an encoding state of higher dimension.  In this case, the encoding states are orthogonal, so Alice could simply measure them in an appropriate basis to determine the angle they encode, and then apply an appropriate unitary operator to her qutrit.  Moreover, one can, in principle, perform an inverse Quantum Fourier Transform to estimate the phase $\theta$.  In the case $\theta=2\pi k/K$,  one can obtain the exact phase with a finite number of anicllas used in the phase estimation.  Below, however, we will use a different strategy, which we can then extend to the case of a general angle.  For a general angle, the protocol becomes either approximate or probabilistic.  

\subsubsection{Sending unknown restricted angle}

Here we discuss the case with two receivers and demonstrate that Alice can send an angle unknown to her, but encoded in a state (with a label $d$).
The case with $N$ receivers will be given in Appendix~\ref{sec:sending-phases-without-measuring-encoding-state}.

We start with the state
\begin{eqnarray}
&&(\alpha^{2} |0\rangle_{a}|v_{0}\rangle_{bc} + \alpha\beta |1\rangle_{a}|v_{1}\rangle_{bc} + \beta^{2}|2\rangle_{a} |v_{2}\rangle_{bc}) \nonumber \\
&&\quad\times\frac{1}{\sqrt{K}} \left( \sum_{j=0}^{K-1} e^{2\pi i kj/K} |j\rangle_{d} \right) .
\end{eqnarray}
Alice has a qutrit $a$, Bob and Charlie have qubits $b$ and $c$, respectively, and the angle $2\pi k/K$ is encoded by a $K$-dimensional qudit $d$, which has been sent to Alice.  Here $|v_{0}\rangle_{bc} = |00\rangle_{bc}$, $|v_{1}\rangle_{bc}= |01\rangle_{bc}+ |10\rangle_{bc}$, and $|v_{2}\rangle_{bc}= |11\rangle$.   Alice does not know what the angle is.  She applies a Controlled-Shift operation to the $ad$ system, with $a$ as the control, which acts as $|j\rangle_{a} |k\rangle_{d} \rightarrow |j\rangle_{a} |k+j\rangle_{d}$, where the addition is modulo $K$.  This results in the following state
\begin{eqnarray}
\frac{1}{\sqrt{K}} \left[ \alpha^{2} |0\rangle_{a} |v_{0}\rangle_{bc} \left( \sum_{j=0}^{K-1} e^{2\pi i kj/K} |j\rangle_{d} \right) \right. \nonumber \\
+ \alpha\beta |1\rangle_{a} |v_{1}\rangle_{bc} \left( \sum_{j=0}^{K-1} e^{2\pi i kj/K} |j+1\rangle_{d} \right)
\nonumber \\
\left. + \beta^{2} |2\rangle_{a} |v_{2}\rangle_{bc} \left( \sum_{j=0}^{K-1} e^{2\pi i kj/K} |j+2\rangle_{d} \right) \right].
\end{eqnarray}
 Using the identity
\begin{equation}
 \sum_{j=0}^{K-1} e^{2\pi i kj/K} |j+l\rangle_{d} = e^{-2\pi i kl/K}  \sum_{j=0}^{K-1} e^{2\pi i kj/K} |j\rangle_{d} ,
\end{equation}
the above state becomes
\begin{eqnarray}
e^{-2\pi i k/K} \left[ (\alpha e^{i\pi k/K})^{2} |0\rangle_{a}|v_{0}\rangle_{bc} + \alpha\beta |1\rangle_{a}|v_{1}\rangle_{bc} \right. \nonumber \\
\left. + (\beta e^{-i\pi k/K} )^{2}|2\rangle_{a} |v_{2}\rangle_{bc} \right]  
\frac{1}{\sqrt{K}} \left( \sum_{j=0}^{K-1} e^{2\pi i kj/K} |j\rangle_{d} \right) . 
\end{eqnarray}
If Alice now performs the steps to send qubit states to Bob and Charlie, they will both receive $\alpha e^{i\pi k/K}|0\rangle + \beta e^{-i\pi k/K}|1\rangle$.  Note that the state that encoded the angle is not affected by this procedure, and it can be sent to another Alice, who can use it to transmit the encoded angle to an additional Bob and Charlie.

\subsubsection{Sending unknown general angle}

Now let us now consider  a general angle $\theta$.  We begin with the state
\begin{eqnarray}
&& (\alpha^{2} |0\rangle_{a}|v_{0}\rangle_{bc} + \alpha\beta |1\rangle_{a}|v_{1}\rangle_{bc} + \beta^{2}|2\rangle_{a} |v_{2}\rangle_{bc}) \nonumber \\
&&\quad \times\frac{1}{\sqrt{K}} \left( \sum_{k=0}^{K-1} e^{ik\theta} |k\rangle_{d} \right) .
\end{eqnarray}
Now Alice applies the Controlled-Shift operation as before.  Suppose that she measures her qutrit in the $\{ |u_{j}\rangle\, |\, j=0,1,2\}$ basis (see Eq.\ (\ref{basis})) and obtains $|u_{0}\rangle$ (if she obtains either of the other two basis elements, Bob and Charlie apply their correction operations to remove unwanted phase factors).    
The resulting $bcd$ state is 
\begin{equation}
\label{gen-angle}
|w_{0}\rangle_{bc} |0\rangle_{d} + |w_{1}\rangle_{bc} |1\rangle_{d} + |w_{2}\rangle_{bc} \times
\frac{1}{\sqrt{K}} \Big(  \sum_{k=2}^{K-1} e^{ik \theta } |k\rangle_{d} \Big)  .
\end{equation} 
Here, we have
\begin{eqnarray}
|w_{0}\rangle_{bc} & = & \alpha^{2} |v_{0}\rangle_{bc} + \alpha\beta e^{i(K-1)\theta} |v_{1}\rangle_{bc} \nonumber \\
&& + \beta^{2} e^{i(K-2)\theta} |v_{2}\rangle_{bc}  \\
|w_{1}\rangle_{bc} & = & \alpha^{2} e^{i\theta} |v_{0}\rangle_{bc} + \alpha\beta |v_{1}\rangle_{bc} \nonumber \\
&& + \beta^{2} e^{i(K-1)\theta} |v_{2}\rangle_{bc}  \\
|w_{2}\rangle_{bc} & = & \alpha^{2} |v_{0}\rangle_{bc} + \alpha\beta e^{-i\theta} |v_{1}\rangle_{bc} \nonumber \\
&& + \beta^{2} e^{-2i\theta} |v_{2}\rangle_{bc}\\
&=&(\alpha|0\rangle_b+\beta e^{-i\theta}|1\rangle_b)\otimes (\alpha|0\rangle_c+\beta e^{-i\theta}|1\rangle_c).\nonumber
\end{eqnarray}
At this point, if Alice wants to adopt a probabilistic protocol, the most obvious thing for her to do is to measure the $d$ system in the basis $\{ |k\rangle_{d}\, |\, k=0,1,\ldots K-1\}$.  If she obtains $k=0$ or 1, $|w_0\rangle$ and $|w_1\rangle$ are entangled between Bob and Charlie and can not be corrected locally to the desired broadcast states.  If, instead, she obtains anything except $k=0,1$, the protocol has succeeded, {\it i.e.}, Bob and Charlie obtain the state $|w_2\rangle_{bc}$, and this happens with a probability of $(K-2)/K$, independent of $\theta$.  This procedure, however, destroys any information about the phase remaining in the $d$ state.  
We discuss improvements that allow us to reuse the encoding state to some extent in Appendix~\ref{sec:sending-phases-without-measuring-encoding-state}.

\subsection{Authentication}
Sequences of symbols that disagree in every place could find application in authentication protocols \cite{authentication}.  Alice can authenticate herself to Bob by providing elements of her sequence to him, and he can check that what Alice sent disagrees with his sequence in every place.  The use of checking sequences for elements that were not sent, that is the sequence of states sent and the sequence of measurement results disagree, has been used in digital signature schemes \cite{andersson}.  
If there are more than two parties, then disagreeing sequences can be used to anonymously send messages or to authenticate votes.  
Suppose we have three parties, Alice, Bob, and Charlie, and they possess sequences of four symbols that disagree in all of their places.  
Bob wants to send a message to Alice but does not want her to know from whom it came.  
He attaches part of his sequence to the message and sends it to Alice.  
Alice can then verify that the message came from Bob or Charlie, but not which one.  
In a voting scenario, each voter will have a sequence, and the authority, who counts the votes, will as well.  
Each voter sends their vote accompanied by their sequence.  
The authority can check that each sequence disagrees with his in all places, which authenticates the votes, and he will not know which sequence came from which voter.
One quantum method of generating such sequences was provided, for example, by Cabello~\cite{cabello}, using a super-singlet quantum state.

Here we will show how three random sequences of three symbols can be generated.  
These sequences have the property that the sequences held by Bob and Charlie disagree in every place with the sequence held by Alice, but they do not necessarily disagree with each other.  
It is also the case that Bob does not know Charlie's sequence and vice versa.  
The basis of this procedure is provided by the trine and anti-trine states of a qubit.
Define the trine states for a qubit to be
\begin{eqnarray}
\label{eq:3states}
|\psi_{0}\rangle & = & \frac{1}{\sqrt{2}}(|0\rangle + |1\rangle ) \nonumber \\
|\psi_{1}\rangle & = & \frac{1}{\sqrt{2}} ( e^{2\pi i /3}|0\rangle + e^{-2\pi i /3} |1\rangle ) \nonumber \\
|\psi_{2}\rangle & = & \frac{1}{\sqrt{2}} ( e^{-2\pi i /3}|0\rangle + e^{2\pi i /3} |1\rangle ) ,
\end{eqnarray}
and, similarly, the anti-trine states to be
\begin{eqnarray}
\label{eq:anti3states}
|\bar{\psi}_{0}\rangle & = & \frac{1}{\sqrt{2}}(|0\rangle - |1\rangle ) \nonumber \\
|\bar{\psi}_{1}\rangle & = & \frac{1}{\sqrt{2}} ( e^{2\pi i /3}|0\rangle - e^{-2\pi i /3} |1\rangle ) \nonumber \\
|\bar{\psi}_{2}\rangle & = & \frac{1}{\sqrt{2}} ( e^{-2\pi i /3}|0\rangle - e^{2\pi i /3} |1\rangle ) .
\end{eqnarray}
This is a rotated version of the usual trine and anti-trine states, but these are the ones that are most useful in the kind of procedure we have been discussing.  Note that 
\begin{align}
\langle \psi_{j}|\psi_{k}\rangle &=  \langle\bar{\psi}_{j}|\bar{\psi}_{k}\rangle 
= -\frac{1}{2} \quad (j\neq k)  , \\
&\langle \bar{\psi}_{j}|\psi_{j}\rangle = 0 . 
\end{align}
We now consider the POVM based on the anti-trine states with measurement operators given by 
\begin{equation}
A_{j}=\sqrt{\frac{2}{3}} |\bar{\psi}_{j}\rangle\langle\bar{\psi}_{j}| , \quad \sum_{j=0}^{2} A_{j}^{\dagger}A_{j}=I , 
\end{equation}
and the probability of obtaining result $j$ in the state $\rho$ is Tr$(\rho A^{\dagger}_{j}A_{j})$.  Bob and Charlie will measure their qubits using this POVM and use the results to generate their sequences. 

To generate the sequences, Alice sends a qubit in one of the trine states to Bob and Charlie using the original broadcast procedure.  Bob and Charlie measure their qubits using the anti-trine POVM and use their measurement results as the elements of their sequences.  Both Bob and Charlie will obtain results different from what Alice sent.  For example, if Alice sent $|\psi_{0}\rangle$, then Bob can obtain $1$ or $2$, and Charlie can obtain $1$ or $2$.  Bob does not know Charlie's result and Charlie does not know Bob's, and their results can be the same or different. (The probability of being the same decays exponentially with the size of the sequence.)  However, their results will differ from what Alice sent. 
Thus, Bob and Charlie can separately establish a sequence that can be used for  authentication with Alice. We provide one example in Table~\ref{table:authentication}.

\vspace{10pt}
\begin{table}
\begin{tabular}{|c||c|c|c|c|c|c|}
    \hline
         Slot & 1 & 2 & 3 & 4 & 5 & 6  \\ \hline
         State Alice broadcasts
         & $|\psi_1\rangle$
         & $|\psi_2\rangle$ 
         & $|\psi_1\rangle$ 
         & $|\psi_0\rangle$ 
         & $|\psi_1\rangle$ 
         & $|\psi_2\rangle$  \\ 
         Bob's outcome 
         & $|\bar{\psi}_2\rangle$
         & $|\bar{\psi}_1\rangle$
         & $|\bar{\psi}_0\rangle$
         & $|\bar{\psi}_2\rangle$
         & $|\bar{\psi}_2\rangle$
         & $|\bar{\psi}_0\rangle$\\
        Charlie's outcome 
         & $|\bar{\psi}_0\rangle$
         & $|\bar{\psi}_1\rangle$
         & $|\bar{\psi}_2\rangle$
         & $|\bar{\psi}_2\rangle$
         & $|\bar{\psi}_0\rangle$
         & $|\bar{\psi}_0\rangle$\\\hline
\end{tabular}
\caption{\label{table:authentication} An example of how to generate keys for authentication from the broadcast states.}
\end{table}

If there are two Alices, as in Sec.~\ref{sec:twosenders}, the qubits received by Bob and Charlie will depend on the angles chosen by both of them.  In particular, the angle will be the sum of the angles chosen by Alice 1 and Alice 2.  In order to authenticate Bob or Charlie, the Alices would have to cooperate. 

{We note that in the case with two receivers and single sender, Alice simply needs to set $\alpha=\beta=1/\sqrt{2}$ and $\theta=0$ in the BBP, then one of the trine states is generated by random phases that come from her measurement.
One can  also consider the  direct broadcast of the trine states to $N$ receivers, where Alice would make use of $\alpha$, $\beta$, and $U(\theta)$, and communicate her measurement result to let receivers correct the unwanted phases. 
}

\subsection{Three-state QKD}
Phoenix, Barnett and Chefles (PBC)~\cite{phoenix} proposed a three-state QKD protocol by using the three states in Eq.~(\ref{eq:3states}). The sender, Alice will randomly choose one of the three states  to send to the receiver, Bob, who then randomly chooses one of the three measurement bases 
\begin{equation}
\{|\psi_j\rangle, |\bar{\psi}_j\rangle\} \quad  (j=0,1,2)
\end{equation}
to measure the received qubit.  We first describe their protocol (Alice and a single Bob) and it will then be obvious that our broadcasting protocol allows Alice to establish separate secret keys with two Bobs. 

\begin{figure*}
\includegraphics[width=0.4\linewidth]{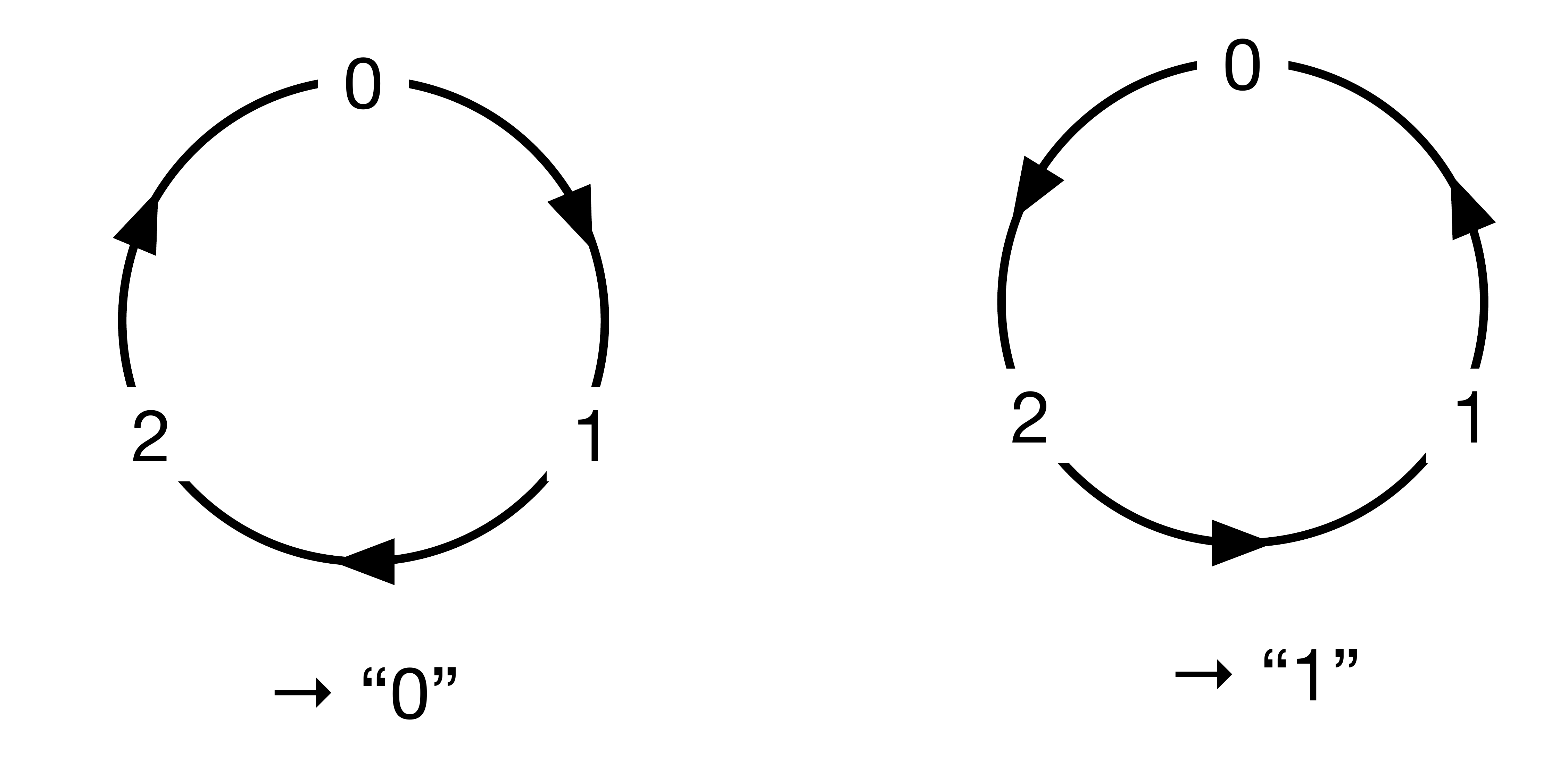}
\caption{Encoding of secret keys in PBC QKD. The numbers within each circle represent the label $j$ of a trine state $|\psi_j \rangle$ (Eq.~\eqref{eq:3states}). One reads the pair of labels $j$: the label of the sent state (which either Alice knows herself or Bob infers from his measurement), $k$: the label of the state not sent (as announced by Alice), and see if the arrow ``$j \rightarrow k $" flows anti-clockwise or clockwise, which encodes ``0" and ``1," respectively. }
\label{fig:threeQKD}
\end{figure*}

First, if Bob's measurement outcome corresponds to $|\psi_j\rangle$, then he announces the failure of this round. Otherwise, Alice announces one of the states in  Eq.~(\ref{eq:3states}) that she did not send. Suppose Bob measures in the basis labeled by $j=1$ and Alice announces that she did not send $j=1$, then there is nothing Bob can infer, as his measurement outcome $|\bar{\psi}_1\rangle$ basically confirms this. The useful case is that the state Alice announces has a different label than Bob's measurement basis label. For example, if she announces that she did not send $j=0$, then Bob can infer that she must have sent $|\psi_{j=2}\rangle$ and we can label this event as (sent label, state not sent)=(2,0). Then using the rule that from the first index to the second index, if there is a +1 hop (mod 3), then it is assigned a 0. On the other hand, if Alice announces that she did not send $j=2$, then Bob can infer the state Alice sent has the label $j=0$, then the event is labeled as (0,2), which requires +2 to get from 0 to 2 modulo 3, and hence this corresponds to a 1. 
We illustrate the convention of assigning whether 0 or 1 in  Fig.~\ref{fig:threeQKD}.
In Table~\ref{tab:example-QKD}, we provide an example of the PBC protocol.

\begin{table*}
    \centering
    \begin{tabular}{|c||c|c|c|c|c|c|c|c|c|c|}
    \hline
         Time & 1 & 2 & 3 & 4 & 5 & 6 & 7 & 8 & 9 & 10 \\ \hline
         State Alice broadcasts: $|\psi_j \rangle$
         & $|\psi_1\rangle$
         & $|\psi_2\rangle$ 
         & $|\psi_1\rangle$ 
         & $|\psi_0\rangle$ 
         & $|\psi_1\rangle$ 
         & $|\psi_2\rangle$ 
         & $|\psi_0\rangle$ 
         & $|\psi_0\rangle$ 
         & $|\psi_2\rangle$ 
         & $|\psi_1\rangle$ \\ 
         Bob's measurement $\{|\psi_\ell \rangle,|\bar{\psi}_\ell \rangle\}$ &  1&1&0&2&1&1&1&2&0&1  \\ 
         Outcome 
         & $|\psi_1\rangle$
         & $|\bar{\psi}_1\rangle$
         & $|\psi_0\rangle$
         & $|\bar{\psi}_2\rangle$
         & $|\psi_1\rangle$
         & $|\bar{\psi}_1\rangle$
         & $|\psi_1\rangle$
         & $|\bar{\psi}_2\rangle$
         & $|\bar{\psi}_0\rangle$
         & $|\psi_1\rangle$\\ 
         Alice announces not ($k$) 
         &  
         & 0
         &
         & 1
         &
         & 1
         &
         & 2
         & 1
         & \\ 
         Bob successfully infers $|\psi_j \rangle$ 
         &
         & \checkmark
         &
         & \checkmark
         &
         & No
         &
         & No
         & \checkmark
         & \\ 
         $j\rightarrow k$ 
         &
         & $2\rightarrow 0$
         &
         & $0\rightarrow 1$
         &
         &
         &
         &
         & $2\rightarrow 1$
         & \\ 
         Shared bit 
         &
         &0
         &
         &0
         &&&&
         & 1
         & \\ \hline
    \end{tabular}
    \caption{An example of the PBC 3-state QKD protocol. In the original design~\cite{phoenix}, Bob performs a randomly chosen 2-outcome projective measurement and this can be replaced by a POVM defined by the anti-trine states.  }
    \label{tab:example-QKD}
\end{table*}

Doing the above for a long sequence $k$ of repetitions, then one can establish a secret key of an expected length $k/3$. Note that Bob can use a POVM defined by $\{\frac{2}{3} |\bar{\psi}_j\rangle\langle \bar{\psi}_j|, j=0,1,2\}$ then one gains a factor of 3/2 and the expected key length is $k/2$.
In our protocol, {\it Alice can broadcast to two Bobs randomly but the same state.}
{Namely, we use the broadcast state $|\Psi^{(1,2)}\rangle$ with $\alpha=\beta=\frac{1}{\sqrt{2}}$ and $\theta=0$, then two Bobs have one of the trine states. }
However, the two Bobs may have different choices of projective measurement bases  or in the case of POVM they may have different POVM outcomes.  From the PBC protocol, it is obvious that Alice and each Bob share a different secret key. (The two Bobs may also share a smaller subset of the key.)

{
Obviously, we can have more than two receivers, in which case we use $|\Psi^{(1,N)}\rangle$ to distribute the same state to $N$ receivers.
Instead of random phases arising from Alice's measurement, Alice can make use of either $U(\theta)$ or the preset parameters $\alpha, \beta$ to choose which trine state to be sent, and let receivers correct the unwanted phases from the measurement $e^{2 \pi n j/(N+1)}$ ($n=0,..,N$).
}

\section{Broadcasting a distributed stabilizer state}
\label{sec:broadcasting-a-distributed-stabilizer-state}
In this section, we discuss different ways of broadcasting a distributed stabilizer state $|\psi\rangle$, where $S|\psi\rangle=|\psi\rangle$ for any $S$  being a product of Pauli operators of a group $G$ that does not contain $-I$. 
 The sender(s) has the choice of distributing the state or aborting the mission. We will mostly focus on graph states as an illustration, to which cluster states are special cases. Moreover, all stabilizer states are locally equivalent to graph states~\cite{VanDenNest2005}.  A graphs state is defined in terms of a graph that consists of a set of vertices $V$ and a set of edges $E$ such that the stabilizer is of the form
 $K_v = X_v\bigotimes_{\langle u,v\rangle \in E} Z_u$ with $v \in V$. (Note that we have used the symbol $K$ for graph-state stabilizer instead of $S$, as this is the convention in the literature.)
Since $K_v = (\prod_{\langle u,v\rangle \in E} CZ_{v,u})^{\dagger} X_v  (\prod_{\langle u,v\rangle \in E} CZ_{v,u})$, the graph state $|\psi_G\rangle$ is given by 
\begin{align}\label{eq:graph-state}
    |\psi_G\rangle & =  U_{CZ }|+\rangle^{\otimes |V|}\ , \quad  
    U_{CZ}=\prod_{\langle u,v\rangle \in E} CZ_{u,v} 
    \ . 
\end{align}
One of the approaches we discuss can be used for states that are not stabilizer states.  

\subsection{Distribution of graph states (I) }
\label{sec:distribution-of-graph-I}
The approach in this section works for graph states and other non-stabilizer states, such as hypergraph states and other states for which the key entangling part consists of only phases. 

Let us first consider the simplest example of distributing a graph state, extending the setup in the Introduction.
Namely, we have Alice (sender) and two Bobs (receivers), and in the end, the two receivers will share two qubits which comprise two-qubit graph state ($V=\{v_1,v_2\}$). 
We consider a state to be shared by the three players, which is proportional to
\begin{align}
    |0\rangle_a |0,0\rangle_{b_1,b_2} 
    + |1\rangle_a ( |0,1\rangle + | 1,0 \rangle)_{b_1,b_2}
    + |2\rangle_a |1,1\rangle_{b_1,b_2} \ .
\end{align}
This state is a special broadcast state with $\alpha=\beta=\frac{1}{\sqrt{2}}$ and can be created by Alice locally. 
For the graph state distribution, Alice applies a phase gate so that the state is now proportional to
\begin{align}
    |0\rangle_a |0,0\rangle_{b_1,b_2} 
    + |1\rangle_a ( |0,1\rangle + | 1,0 \rangle)_{b_1,b_2}
     -|2\rangle_a |1,1\rangle_{b_1,b_2}.
\end{align}
She distributes the corresponding qubits to the two Bobs respectively. Then Alice measures her qutrits in the basis $\{|\tilde{j}\rangle \}$,
\begin{align}
    |\tilde{j}\rangle = \frac{1}{\sqrt{3}}\sum_{k =0,1,2} e^{2 \pi i jk/3} |k \rangle \quad (j=0,1,2) \ .  
\end{align}
The resulting state for two receivers is
\begin{align}
&
\left( 
|0,0\rangle + e^{- 2 \pi i j/3} (|0,1\rangle + | 1,0\rangle) - e^{- 4 \pi i j/3} | 1,1 \rangle 
\right)_{b_1,b_2} \nonumber \\
&=
CZ_{b_1,b_2} \left( |0 \rangle + e^{-2 \pi i j/3} |1\rangle\right)_{b_1} 
\left( |0 \rangle + e^{-2 \pi i j/3} |1\rangle \right)_{b_2} \ .
\end{align}
We note that the extra phase $e^{-2 \pi i j/3}$ commutes with the $CZ$ operator, so it is locally correctable by each receiver if Alice informs them of the measured value of $j$, as in the previous broadcasting case. 

If Alice decides to abort the protocol instead, she should change her measurement basis to $\{|0\rangle, \, |1\rangle, \, |2\rangle  \}$. This may leave behind an entangled state shared among Bobs.

Generalization to graphs with $N$ vertices is straightforward.
In order to distribute a graph state with a general graph connectivity, the $CZ$ gate should be applied locally by Alice before the distribution of physical qubits.
In this scenario, we prepare
\begin{equation}
|\Psi^{(1,N)}_{CZ}\rangle = \sum_{k=0}^{N} 
\left( \begin{array} {c} N \\ k \end{array}\right)^{1/2}
|k\rangle_{a} 
U_{CZ}
|k;N-k\rangle ,
\end{equation}
where $U_{CZ}$ is the product of $CZ$ gates that generates the graph state with the desired connectivity.
After the measurement by Alice (assuming the result is $j$), receivers Bobs obtain the state
\begin{align}
\! \! |\Psi^{(1,N)}_{CZ}\rangle \xrightarrow{\text{measure}}
&\sum_{k=0}^{N} 
\left( \begin{array} {c} N \\ k \end{array}\right)^{1/2}
e^{-2 \pi i j k  / N} 
U_{CZ}
|k;N-k\rangle  \nonumber \\
&=
U_{CZ} \left(
|0\rangle + e^{-2 \pi i j  / N} |1\rangle
\right)^{\otimes N}.
\end{align}
 The resulting state is the graph state \eqref{eq:graph-state}, up to a correctable phase  $e^{-2 \pi i j  / N}$.

In general, any state that can be generated by applying a multi-qubit phase gate (such as a product of $CZ$ gates or $CCZ$ gates) on product states can be distributed using this strategy.
This class of quantum state includes graph states~\cite{Hein}, hypergraph states~\cite{Rossi}, and other short-range entangled states including symmetry-protected topological (SPT)-ordered states~\cite{Chen, Yoshida}, where the phases in the last case come from the cocycles associated with the symmetric phase.

We can generalize the result in Eq.~\eqref{eq:general-distributed-state} so we have multiple senders who add phases after distribution. 
We write the multi-qubit phase gate as $U_{\text{phase}}$, then using 
\begin{align}
    &|\Psi^{(M,N)}_{\text{phase}}\rangle \nonumber \\ 
    &= \sum_{k=0}^{N} \alpha^{k} \beta^{N-k} \left( \begin{array} {c} N \\ k \end{array}\right)^{1/2}
\prod_{j=1}^{M}  |k\rangle_{aj}  U_{\text{phase}} |k;N-k\rangle 
\end{align}
and letting Alices apply the additional phase gate, the following entangled state is distributed among receivers:
\begin{widetext}
\begin{align}
 |\Psi^{(M,N)}_{\text{phase}} \rangle 
 & \xrightarrow[\text{measure}]{U_j(\theta_j)^{\otimes M}}
U_{\text{phase}} \prod_{l=1}^{N}\left( \alpha e^{i\sum_{j=1}^{M} \theta_{j}} e^{-2\pi i (\sum_{j=1}^{M} n_{j})/(N+1)} |0\rangle_{l} + \beta e^{-i\sum_{j=1}^{M} \theta_{j}} |1\rangle_{l} \right)  \\ 
 & \xrightarrow{\text{correct}} 
U_{\text{phase}} \prod_{l=1}^{N}\left( \alpha e^{i\sum_{j=1}^{M} \theta_{j}}  |0\rangle_{l} + \beta e^{-i\sum_{j=1}^{M} \theta_{j}} |1\rangle_{l} \right) .
\end{align}
\end{widetext}

\subsection{Broadcasting a stabilizer state} \label{sec:broadcasting-stabilizer}
Here we describe another protocol to `broadcast' a single stabilizer state distributed among multiple receivers. 
For simplicity, we consider an $N$-qubit stabilizer state, for which there are $N$ independent stabilizer generators $\{S_k\}$. (One can consider fewer stabilizer generators to encode a distributed logical state.) In this case, Alice will need $2N$ qubits, where the first $N$ qubits, all initilaized  in $|+\rangle$, serve as a control set that involves the action of  control-$S_k$'s to the corresponding qubits in the second group,  whose qubits are conveniently initialized all in $|+\rangle$ as well. Given that each $S_k$ is a product of single Pauli operators, the control-$S_k$ gate is decomposed into a product of two-qubit control-Pauli gates. Then Alice sends the qubits in the second group to the desired receivers, Bobs.

The resultant state before any measurement by Alice can be rewritten as
\begin{align}
  |\psi\rangle = \prod_{k=1}^N \frac{1}{\sqrt{2}}\big( |0\rangle_k \otimes I + |1\rangle_k \otimes  S_{k'}\big) \prod_{k'=1}^N |+\rangle_{k'}.
\end{align}

If Alice intends to distribute the stabilizer state, then she performs measurement on her remaining $N$ qubits in the $X$ basis. The outcomes represent the action $(I\pm S_k)/2$ is performed or equivalently the value of the stabilizer operators $S_k=\pm 1$. She can find out the correcting Pauli product and inform the appropriate parties to make their correction operation.

On the other hand, if she intends to abort the mission, she simply measures all her qubits in the $Z$ basis. The measurement outcomes represent whether the corresponding $S_k$ operator was applied or not, and regardless of the outcomes, all Bobs then share a product state.

\subsubsection{Distribution of graph states (II)}
\begin{figure*}
    \centering
    \includegraphics[width=0.7\linewidth]{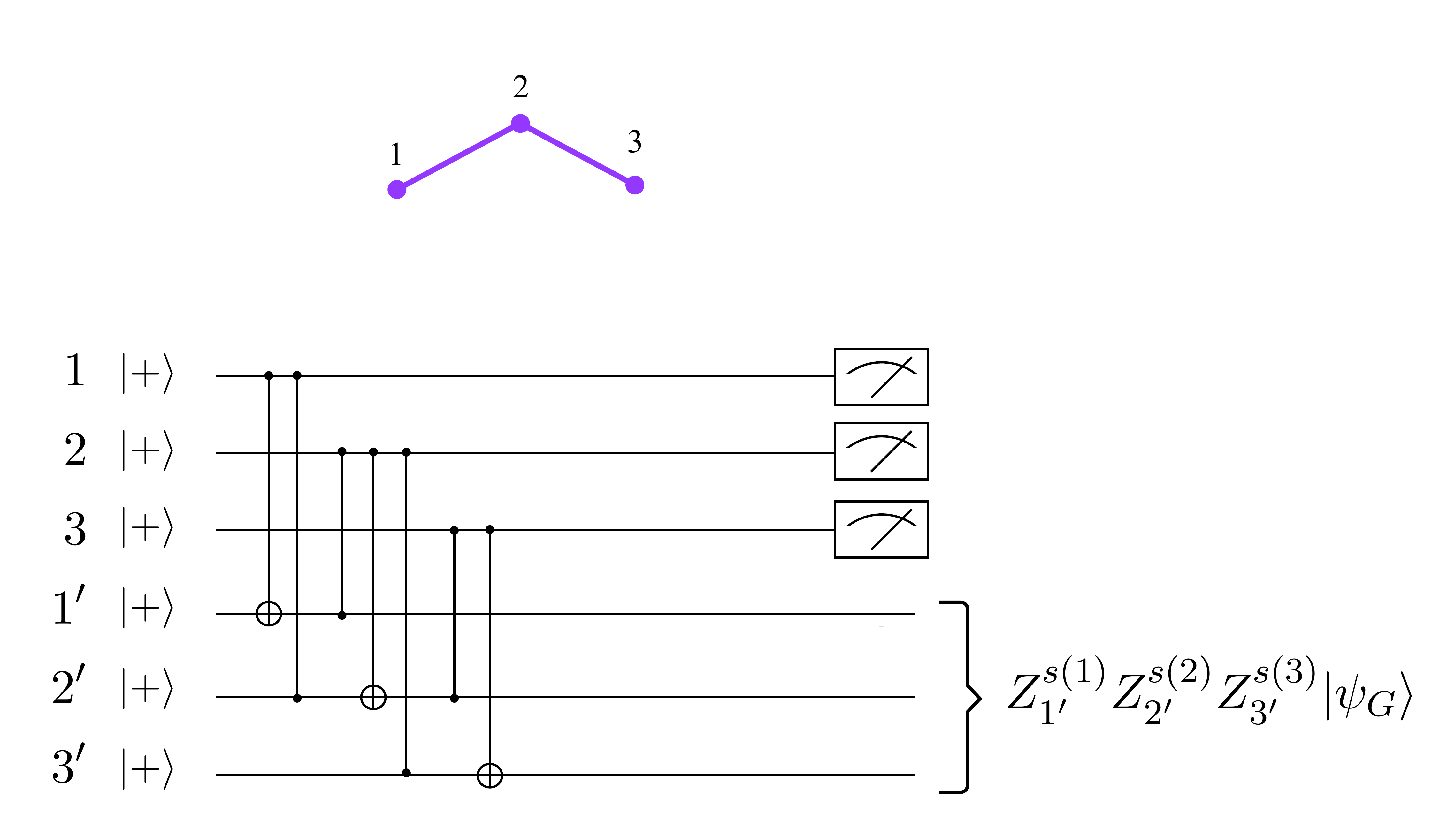}
    \caption{An example of graph state distribution. 
    The graph in this example consists of three vertices, $v \in \{1,2,3\}$, and the graph state is distributed. 
    The graph state $|\psi_G\rangle$ is stabilized by $X_1Z_2$, $Z_1X_2Z_3$, and $Z_2X_3$. 
    }
    \label{fig:distribution-of-graph-states}
\end{figure*}

In the remainder of this section, we specialize to graph states, as a special case of the general stabilizer states.
We consider qubits initialized as the 2$|V|$ product of $|+\rangle$, {\it i.e.}, $\prod_{v \in V} | + \rangle_v \prod_{v' \in V} | + \rangle_{v'}$, where the qubit $v$ will be associated with the stabilizer operator $K_{v'}= X_{v'} \prod_{\langle u',v'\rangle \in E} Z_{u'}$ and the symbol $'$ denotes qubits that will be distributed to Bobs. 
After applying all the control-$K_{v'}$ gates (which can be decomposed into a product of CX and CZ gates), 
\begin{align} \label{eq:broadcasting-state-for-graph-state}
    |\psi \rangle
    = & \prod_{v \in V} \frac{1}{\sqrt{2}} \Big( |0\rangle_v \otimes I + |1\rangle_v \otimes X_{v'} \prod_{\langle u',v'\rangle \in E} Z_{u'}  \Big) \nonumber\\ 
    & \times \prod_{w' \in V} | + \rangle_{w'} \ . 
\end{align}
Denoting the outcome of the Pauli-$X$ measurement for the $v$-qubit of Alice being $s(v) \in \{0,1\}$, the resulting state after the measurement is given by
\begin{align}
\label{eq:projector}
    \prod_{v\in V}
    \frac{1}{2} \Big( 1+ (-1)^{s(v)} X_{v'} \prod_{\langle u',v'\rangle \in E} Z_{u'} \Big) 
    \prod_{w' \in V}
    |+\rangle_{w'},
\end{align}
and this state is a  graph state shared  by Bobs up to by-product operators,
\begin{align}
    \Big(
    \prod_{v \in V} Z_v^{s(v)}\Big)|\psi_G\rangle_{B},
\end{align}
where $|\psi_G\rangle_{B}$ denotes the graph state with $s(v)=0$. 
Note that the operator acting on the product state in Eq.~(\ref{eq:projector}) is a projector.
If the measurement outcome is ideal (i.e., $s(v)=0$ for all $v \in V$), Bobs obtain the state projected to the desired subspace with $K_v=1$ for all $v \in V$, which is the graph state $|\psi_G\rangle$.
If there appear outcomes $s(v)=1$, then one can apply the Pauli-$Z$ operator at corresponding $v'$ to correct the eigenvalue of the stabilizer. 
In Fig.~\ref{fig:distribution-of-graph-states}, we give an example of a graph state whose graph consists of three vertices.

\subsubsection{Rotated graph states}

The previous discussion on sending phases motivates us to devise a method to teleport phase gates in the setup with the graph states.
If one wishes to encode a phase $\theta$ acting on a vertex $v$ held by a Bob, we can, for example, add an additional gate teleportation to the graph state prepared with the procedure above.  
Let us say Alice wants to send a phase gate $e^{i \theta Z}$ to the vertex $u \in V$. 
Then Alice prepares another supplemental qubit in $|+\rangle_{u_A}$ and couples it with $u_{B}$ via $CZ_{u_A,u_{B}}$.
After distributing all the Bobs' qubits, Alice measures her supplemental qubit with the basis \begin{align}
    \mathcal{M}_\theta=\left\{ 
    e^{-i \theta X} |s\rangle  \ \middle| \ s=0,1 \right\} \ .
\end{align}
Then by virtue of the gate teleportation, the unitary gate $(Z)^s e^{i \theta Z}$ is induced at $u_{B}$, which acts on the graph state. 
One can easily generalize this procedure with a different phase gate for each vertex.
In the end, the following state is distributed:
\begin{align} \label{eq:graph-state-with-phase}
    \left( \prod_{v \in V} e^{i \theta_v Z_v } \right) |\psi_G\rangle,
\end{align}
up to $Z$ by-product operators that depend on Alice's measurement outcomes.
We elaborate more details of our scheme in Appendix~\ref{sec:mbqc-detail}.

\section{Applications of distributed stabilizer states}
\label{sec:applications-of-distributed-stabilizer-states}
\subsection{Distribution of GHZ state}

A distributed GHZ state has applications such as quantum secret sharing \cite{QSS}.
The GHZ state can be obtained from a star-graph state and applying Hadamard gate to every qubit but the one at the center of the star \cite{Hein}. 
For example, one can assign Alice to the central qubit, and other qubits are possessed by Bobs. 
After that, all Bob apply Hadamard, then Bobs and Alice share a GHZ state.
In Appendix \ref{sec:GHZ-alternative}, we present an alternative way to distribute a GHZ state.

\subsection{Distributed MBQC with $X$ alone by receivers}

Below, we describe how the phase rotation gate teleported from Alice can be used by Bob to perform the measurement-based quantum computation (MBQC)~\cite{MBQC-PRL, MBQC-PRA}.
With her ability to send phase rotation gates, Alice and Bob can work together to perform the universal MBQC, for which Alice controls the adaptation of the measurement axis (by the rotation) and Bob always performs measurement in the same basis, $X$.  This allows a universal gate set to be implemented.
We also point out in Appendix~\ref{sec:reduction-of-graph} that Alice can remove a vertex from the graph possessed by Bob by appropriately choosing her measurement basis.

\begin{figure*}
\includegraphics[width=0.7\linewidth]{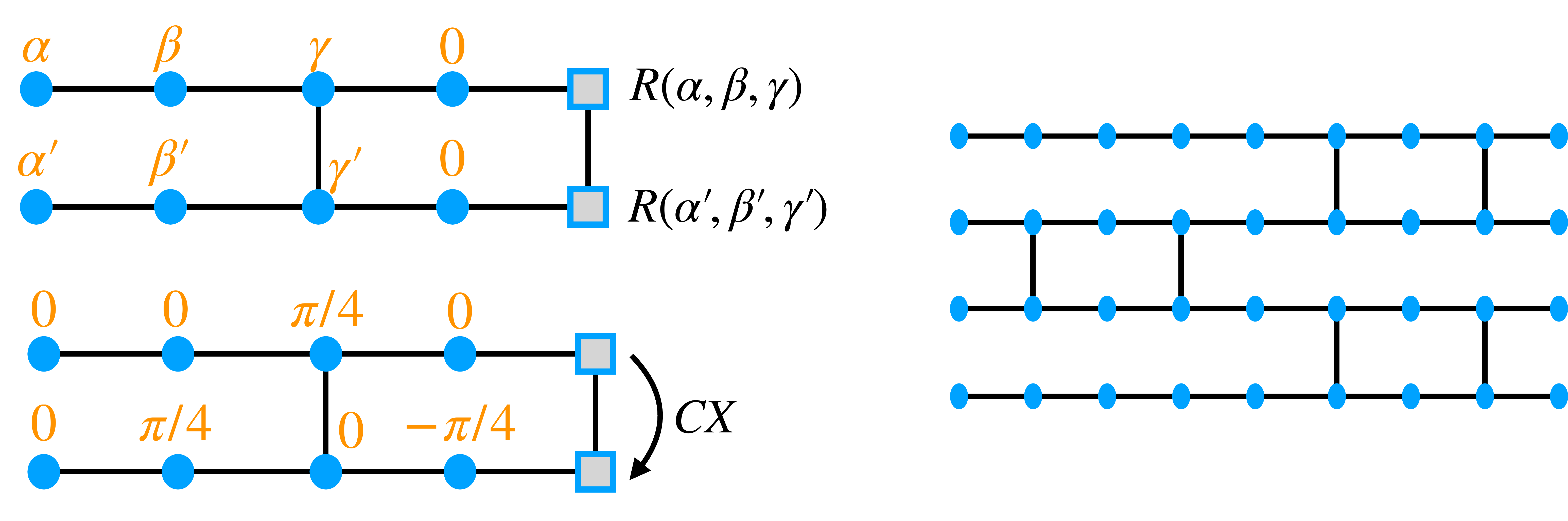}
\caption{(Left top) Measurement pattern to realize the single-qubit $SU(2)$ rotation gate. $R(\alpha, \beta,\gamma)$ represents the Euler rotation. (Left bottom) Measurement pattern to realize the controlled-not gate. 
The information flows from left to right as we sequentially measure qubits in order. 
The number above each ball represents the angle $\theta$ in the measurement basis $\{ e^{i \theta Z} |+ \rangle , e^{i \theta Z} |- \rangle\}$. 
(Right) The brickwork state. 
One tiles up the blocks depicted on the left to concatenate the unitaries sequentially. 
See Ref.~\cite{BlindQC, BQC-npjq} for details.
}
\label{fig:brickwork}
\end{figure*}

\begin{figure*}
\includegraphics[width=0.7\linewidth]{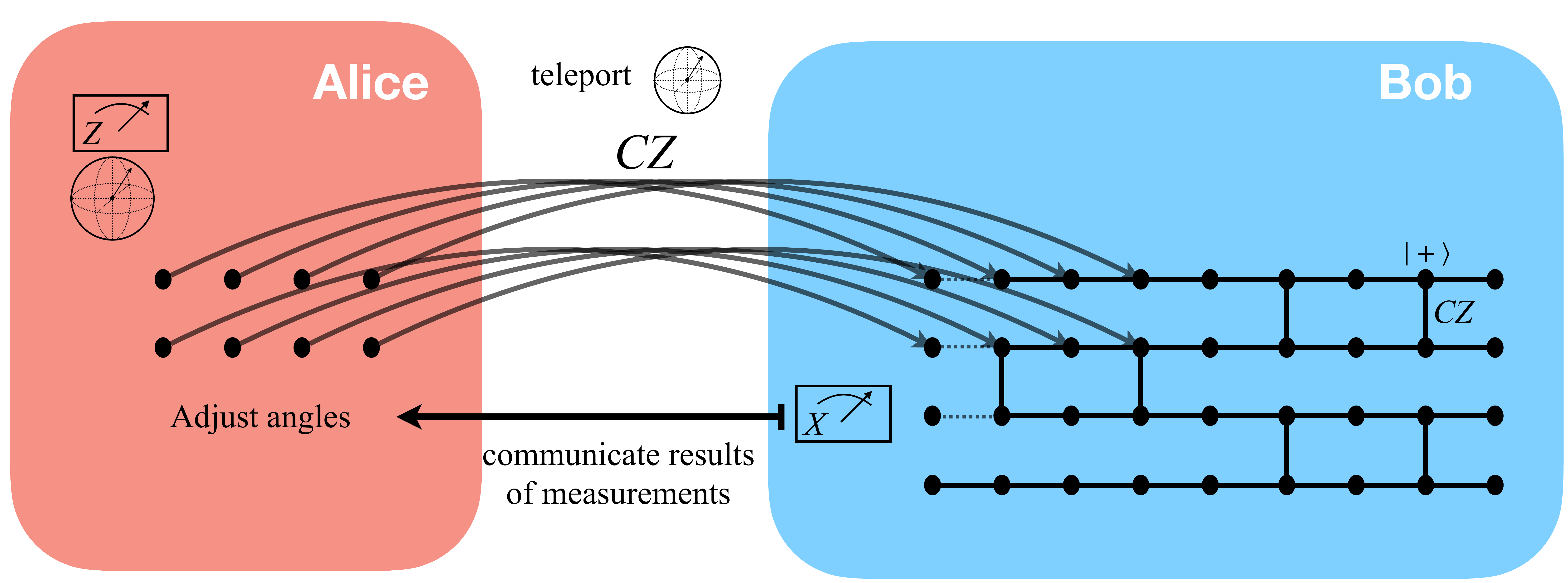}
\caption{
After finishing the phase of distributing the brickwork state to the receiver(s), Alice and Bob can collaborate to perform a universal MBQC. 
}
\label{fig:BQC-Concept}
\end{figure*}

The universal MBQC can be achieved on an appropriate graph state ---the brickwork state--- and measurements with the rotated basis, $B=\{ e^{i \theta Z} |+ \rangle , \, e^{i \theta Z} |- \rangle  \}$ \cite{MBQC-PRL, BlindQC, BQC-npjq}.
Namely, appropriately choosing the angle $\theta$ in the basis $B$, one can achieve single qubit $SU(2)$ rotation gates and the controlled-not gate.  
See Fig.~\ref{fig:brickwork}.
As we discussed in Sec.~\ref{sec:broadcasting-a-distributed-stabilizer-state}, Alice can send a phase gate $e^{i \theta_v Z_v}$ to Bob's qubits, thus the universal MBQC can be performed on the Bob's graph state with the $X$ basis measurement alone by Bob.
Bob needs to inform his measurement outcomes to Alice so she knows the by-product operators from the former measurements acting on his graph state, and she can decide if she needs to teleport $e^{i \theta_v Z_v}$ or $e^{-i \theta_v Z_v}$ to perform desired rotation.
See Fig.~\ref{fig:BQC-Concept}.
We delegate technical details to Appendix~\ref{sec:mbqc-detail}.

\section{Conclusion}
\label{sec:conclusion}

We extended our results for the broadcasting of quantum states to multiple receivers to now include the possibility of multiple senders.  It is possible to add and delete senders from the network, and procedures for doing so were discussed.  We also provided further results on sending states with an unknown phase.  In this case, the phase is provided to the sender encoded in a quantum state.  We showed how the encoding state can be reused, and, for a general phase, how each use degrades the information contained in it.  We went on to give two applications of our broadcasting protocol, including authentication and three-state quantum cryptography.

We have considered the distribution of graph states in two different ways.
The first method generalizes the state that is originally used for broadcasting product states, by supplementing multi-qubit phase gates that entangle the product states. This applies beyond graph states, including hypergraph states and nontrivial, fixed-point symmetry-protected topologically ordered states.
Another way to broadcast a graph state is to implement projection controlled by the sender.
The sender Alice can decide the structure of the graph by choosing her measurement basis appropriately, and she can also send additional phase gates that act on the graph state.
We have also discussed applications of the distribution of the graph state; the distribution of the GHZ state and the remote, distributed universal quantum computation done by the distributed receivers with the $X$ basis measurement alone and by the sender who controls the (adaptive) rotation. The latter is in the same the reverse of the blind quantum computation; here the server is outsourcing computational tasks to third parties, without them having the possibility of knowing the computation.

\acknowledgments
The authors would like to thank Himanshu Gupta for useful comments.   The research of MH, JB, HS, and T-CW was supported by NSF grant FET-2106447. The research of MH and JB was also sponsored by the DEVCOM Army Research Laboratory and was accomplished under Cooperative Agreement Number W911NF-20-2-0097. Additionally, T-CW and HS would like to acknowledge the  support of a Provost's Venture Fund from the  Provost's Office at Stony Brook University.

\appendix

\section{More on sending unknown phases}
\label{sec:sending-phases-without-measuring-encoding-state}

\subsection{Sending unknown restricted angle to $N$ receivers}

Let us start with the state
\begin{align}
|\Psi^{(1,N)}\rangle \otimes \frac{1}{\sqrt{K}} \Big( 
\sum^{K-1}_{j=0} e^{2 \pi i k j /K} | j \rangle_d 
\Big)  .
\end{align}
As we did for the case with $N=2$ (Bob, Charlie) in the main text, Alice applies the Controlled-Shift gate $CX_{a,d}$, which gives us
\begin{align}
&\sum_{\ell=0}^{N} 
\alpha^{\ell} \beta^{N-\ell} \left( \begin{array} {c} N \\ \ell \end{array}\right)^{1/2}
|\ell\rangle_{a} |\ell;N-\ell\rangle \nonumber \\ 
&\qquad \times \frac{1}{\sqrt{K}}\sum^{K-1}_{j=0} e^{2 \pi i j k /K} | j + \ell \rangle_d  . 
\end{align}
By shifting the summed index $j$ by $\ell$, we obtain 
\begin{align}
&\sum_{\ell=0}^{N} 
\big(e^{- 2 \pi i k  /K}
\alpha \big)^{\ell} \beta^{N-\ell} \left( \begin{array} {c} N \\ \ell \end{array}\right)^{1/2}
|\ell\rangle_{a} |\ell;N-\ell\rangle \nonumber \\ 
&\qquad \otimes  \Big( \frac{1}{\sqrt{K}}\sum^{K-1}_{j=0} e^{2 \pi i j k /K} | j  \rangle_d \Big)
\end{align}
Alice measures her qudit in the Fourier basis, then each receiver obtains the following state after the correction of phases induced by Alice's measurement:
\begin{align}
e^{- 2 \pi i k  /K}
\alpha  | 0\rangle + \beta |1 \rangle \ . 
\end{align}

\subsection{Reusing the enconding state with general angle}

In the main text, we discussed methods to send unknown angle $\theta$ encoded in a state $d$ to Bob and Charlie. 
In the case with a general angle, we showed that Alice can measure the encoding state $d$ with the computational basis and with a probability $(K-2)/K$ the procedure succeeds, but the encoding state will be destroyed.
In this section, we discuss alternatives so we can reuse the encoding state. 

A better procedure is the following.  Define
\begin{equation}
Q_{2} = \sum_{k=2}^{K-1} |k\rangle_{d}\langle k| .
\end{equation}
Alice measures $Q_{2}$, which has eigenvalues $0$ and $1$, on the $d$ system, and if she obtains $1$, which she does with probability $(K-2)/K$ the procedure has succeeded, and the resulting $d$ state is $(1/\sqrt{K-2}) \sum_{k=2}^{K-1} e^{ik\theta} |k\rangle_{d}$.  This state does not contain as much information about the phase as the original state, but it still contains quite a bit.  It could be passed on to a second Alice, who can then use it in the same protocol for a different Bob and Charlie. The subsequent success probability will be reduced to $(K-4)/(K-2)$ corresponding to eigenvalue 1 of \begin{equation}
Q_{4} = \sum_{k=4}^{K-1} |k\rangle_{d}\langle k|.
\end{equation}

An alternative is for Alice not to measure the $d$ system, and this will leave Bob and Charlie with approximate versions of the desired state.  In this case the $d$ state can be reused, though it will have been altered somewhat by its first use.  In order to see what Bob and Charlie have, we can form a density matrix from the state in Eq.\ (\ref{gen-angle}) and trace out $c$ and $d$ to obtain Bob's state (Charlie's state is identical).  The result is
\begin{eqnarray}
\rho_{b} & = & \frac{K-2}{K} (\alpha e^{i\theta /2} |0\rangle + \beta e^{-i\theta /2} |1\rangle ) \nonumber \\
& & (\alpha^{\ast} e^{-i\theta /2}\langle 0| + \beta^{\ast} e^{i\theta /2}\langle 1|) \nonumber \\
& & + \frac{1}{K} \left[ 2|\alpha |^{2}|0\rangle\langle 0| + 2|\beta |^{2} |1\rangle\langle 1| \right. \nonumber \\
& &+ \alpha\beta^{\ast} e^{i\theta}(1+e^{-iK\theta}) |0\rangle\langle 1| \nonumber \\
& & \left. + \alpha^{\ast}\beta e^{-i\theta}(1+e^{iK\theta}) |1\rangle\langle 0| \right] 
\end{eqnarray}
Note that the size of the error is of order $1/K$.  

As noted, the state encoding the phase can be reused.  In order to see how much it has been changed by one use, we can calculate the fidelity of the reduced density matrix of the $d$ system after one use, $\rho_{d}$, which can be found by tracing $b$ and $c$ out of Eq.\ (\ref{gen-angle}), to the original encoding state, $|\Phi(\theta , K)\rangle = (1/\sqrt{K})\sum_{k=0}^{K-1} e^{ik\theta}|k\rangle_{d}$.  We find that
\begin{align}
F&=\langle\Phi (\theta,K)|\rho_{d}|\Phi (\theta ,K)\rangle \nonumber \\
 &= 1-4|\beta |^{2}
 (1-\cos K\theta )\frac{1}{K^{2}} (K-2+|\alpha |^{2}) ,
\end{align}
so that the error induced by the first use is of order $1/K$.  If the $d$ state is now sent on to a new Alice, whom we shall call Alice$^{\prime}$, who then uses it to send the angle to Bob$^{\prime}$ and Charlie$^{\prime}$, the resulting $bcb^{\prime}c^{\prime}$ density matrix is of the form
\begin{equation}
\rho_{bcb^{\prime}c^{\prime}} = \frac{K-4}{K} \rho_{prod} + \rho_{noise} ,
\end{equation}
where $\rho_{prod}$ is the density matrix corresponding to the product state $(e^{i\theta /2}\alpha |0\rangle_{bc} + e^{-i\theta /2} \beta |1\rangle_{bc}) (e^{i\theta /2}\alpha |0\rangle_{b^{\prime}c^{\prime}} + e^{-i\theta /2} \beta |1\rangle_{b^{\prime}c^{\prime}})$, and $\rho_{noise}$ is an incoherent superposition of four states and is of size (operator norm) of order $1/K$.  If $\exp (iK\theta )=1$, then $\rho_{noise} = (4/K)\rho_{prod}$.  For $K$ large, this procedure will provide both sets of Bob and Charlie with good approximations of the desired state.  Further uses will degrade the encoding state further, and add more noise to the transmitted states.  A more extensive analysis of the degradation of phase information stored in a quantum state with use is given in \cite{croke}.

\section{Reduction of graph by $Z$ measurements}
\label{sec:reduction-of-graph}

\begin{figure*}
    \centering
    \includegraphics[width=0.4\linewidth]{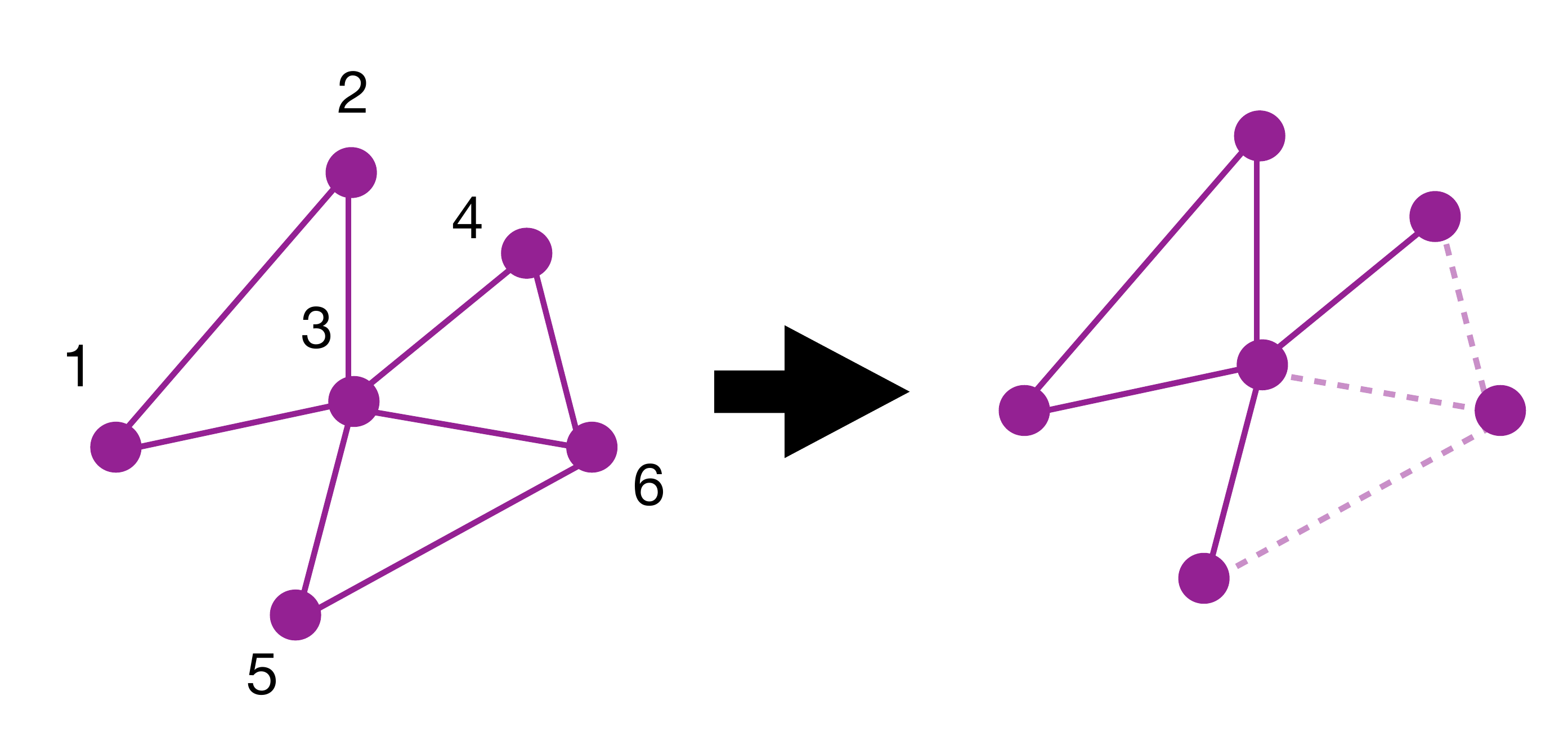}
    \caption{An example of reduction of a graph. $V=\{1,2,3,4,5,6\}$, $E=\{12,13,23,34,35,36,46,56\}$, $V^R=\{1,2,3,4,5\}$, $E^R=\{12,13,23,34,35\}$, $\bar{V}=\{6\}$, $\bar{E}=\{36,56,46\}$.
    }
    \label{fig:graph-reduction}
\end{figure*}

Alice can decide to disconnect vertices from the graph, even after the distribution of qubits to Bob.
(We assign $|+\rangle^{\otimes |V|}$ for Alice and $|0\rangle^{\otimes |V|}$ for Bob instead.)
Suppose the original graph consists of vertices $v \in V$ and edges $e \in E$, but Alice wishes to reduce it to a graph that is made of vertices $V^{R} \subset V$ and edges $E^R \subset E$. 
We denote $\bar{V}=V \backslash V^R$ and $\bar{E}=E \backslash E^R$.
See Fig. \ref{fig:graph-reduction}.

Alice measures her qubits that correspond to $V^R$ with the $X$ basis and those to $\bar{V}$ with the $Z$ basis.
Then Bob obtains
\begin{align}
    &
    \left( \prod_{v \in V^R} \frac{1+ (-1)^{s(v)} K_v  }{2} \prod_{u \in \bar{V}} K_u^{s(u)} \right)
    \prod_{v \in V^R} |0\rangle_v \prod_{u \in \bar{V}} |0\rangle_u  \nonumber \\
    &=
    \left( \prod_{v \in V^R} \frac{1+ (-1)^{s(v)} K_v  }{2}  \right) 
    \prod_{v \in V^R} |0\rangle_v \prod_{u \in \bar{V}} |s(u)\rangle_u \ ,
\end{align}
with $K_v= X_v \prod_{\langle u,v\rangle} Z_u$, which is the graph-state stabilizer operator $S_v$. 
Now we write the stabilizers for the reduced graph as
\begin{align}
    K^{R}_v = X_v \prod_{\langle w,v\rangle \in E^R} Z_w \quad (v \in V^R) \ .
\end{align}
Then defining 
\begin{align}
   \prod_{\substack{ \langle u,v \rangle \in \bar{E} } } (-1)^{s(u)} \equiv  (-1)^{t(v)}  \ 
\end{align}
with $t(v)=0,1$, the resultant state is given by
\begin{align}
    \left( \prod_{v \in V^R} \frac{  1+ (-1)^{s(v)+t(v)}  K^R_v}{2}   \right) 
    \prod_{v \in V^R} |0\rangle_v \prod_{u \in \bar{V}} |s(u)\rangle_u 
    \ . 
\end{align}
This is a graph state with the graph $(V^R,E^R)$ and the product state for the rest of qubits.

\section{Alternative way to distribute GHZ}
\label{sec:GHZ-alternative}

Here we consider the distribution of 
a $N$-qubit Greenberger-Horne-Zeilinger (GHZ)-like state.
We consider a graph state given by vertices $V=\{1,2,...,2N\}$ and edges $E=\{ \langle 1,2\rangle , \langle 2,3\rangle, ... , \langle 2N-1,2N \rangle, \langle 2N,1\rangle \}$.
Namely, this is a 1-dimensional periodic chain labeled by integers.
We use the quantum circuit described in Section~\ref{sec:broadcasting-stabilizer} to generate the broadcasting state eq.~\eqref{eq:broadcasting-state-for-graph-state}.
However this time, the difference is that among the target qubits, the odd part $V_o=\{1,3,...,2N-1\}$ is distributed among receivers, while the even part $V_e=\{2,4,...,2N\}$ is kept by Alice.
Note that Alice also keeps her qubits that controls the stabilizer operators acting on Bobs' qubits, so that she has $|V|+|V_e|(=|V|+|V|/2)$ qubits.
After Alice performs her measurement on her control qubits, due to the projection, Alice and Bobs share the graph state stabilized by
\begin{align}
    & Z_{2n-1} X_{2n} Z_{2n+1} | \psi_{G} \rangle = |\psi_{G} \rangle \label{eq:stabilizer-1}  \\
    & Z_{2n} X_{2n+1} Z_{2n+2} | \psi_{G} \rangle = |\psi_{G} \rangle  \label{eq:stabilizer-2}
\end{align}
up to $\pm 1$ signs.
Now, Alice measures the even part $V_e$ with the $X$ basis.
Because of the new stabilizer conditions from eq.~\eqref{eq:stabilizer-1} becomes,
\begin{align} \label{eq:stabilizer-GHZ}
    Z_{2n-1} X_{2n} Z_{2n+1} = (-1)^{s_{2n}}Z_{2n-1} Z_{2n+1} = 1 
\end{align}
and the global symmetry  $X_{1}...X_{2n+1}...X_{2N+1}=1$ derived from the condition eq.~\eqref{eq:stabilizer-2} remains, then the state shared by receivers is now a GHZ-type state~\cite{GHZ} up to phase flips. 
This technique was given in, for example, Ref.~\cite{Tantivasadakarn}.

\section{Some technical details on distributed remote MBQC} 
\label{sec:mbqc-detail}

The aim of this section is to give readers some details of distributed remote MBQC. 
We will focus on explanation of our method and along the way we give basic ingredients commonly known in the field of MBQC. 
The original idea of using the brickwork state for the blind universal quantum computation was from Ref.~\cite{BlindQC}.
Readers may find it helpful to look at Ref.~\cite{tcw-review, BQC-npjq} for pedagogical introduction to MBQC on the brickwork state.

\subsection{Rotated graph state and measurement}
Consider a graph state depicted in Fig.~\ref{fig:brickwork} (Left).
We label the 10 qubits using an ordered set $V=\{1,2,3,4,5,6,7,8,9,10\}$ from top left, top right, bottom left, to bottom right. 
The pairs $(1,2)$ $(2,3)$ $(3,4)$ $(4,5)$ $(6,7)$ $(7,8)$ $(8,9)$ $(9,10)$ are entangled with the $CZ$ gate in the horizontal direction. 
The pairs $(3,8)$ and $(5,10)$ are entangled with the $CZ$ gate in the vertical direction as well. 
We denote the set of these edges as $E_{bw}$.
The initial information is encoded at $(1,6)$ qubits (the most left pair) as $|\psi\rangle_{1,6}$.
The brickwork state in this setup is written as
\begin{align}
|\psi_{bw}\rangle := \prod_{\langle u,v\rangle \in E_{bw}} CZ_{u,v} \Big( |\psi\rangle_{1,6} \otimes |+\rangle^{\otimes V \backslash \{1,6\} }  \Big)  .
\end{align}

Alice possesses a copy of 10 qubits to teleport $XY$-plane rotations.
We label them by the subscript $A$, while the qubits of the brickwork state possessed by Bob will be labeled by the subscript $B$. 
Consider a state (which is the state after the projection with the stabilizers and corrections by Bob in our broadcast protocol), 
\begin{align}
|\Psi \rangle_{AB} 
= \prod_{v \in V} CZ_{v_A,v_B} 
\Big( | \psi_{bw} \rangle_B \otimes |+\rangle^{\otimes V}_A  \Big) . 
\end{align}
Now consider measurements by Alice with the basis $\{e^{-i \theta X} |0 \rangle , e^{-i \theta X} |1 \rangle \}$. 
Writing the measurement outcome at $v$ as $s_v \in \{0,1\}$ and the angle as $\theta_v $, the post-measurement state is 
\begin{align}
& \Big( \bigotimes_{v \in V}  \langle s_v | e^{i \theta_v X_{v}} \Big)_A  \,|\Psi \rangle_{AB} \nonumber \\ 
&\propto
\Big( \prod_{v \in V } Z^{s_v}_v e^{i \theta_v Z_v}  \Big) | \psi_{bw} \rangle_B  =: |\psi_{bw}({\pmb \theta};{\pmb s}) \rangle 
\end{align}
up to a normalization constant.

Now we move on to consider Bob's measurements with the basis $\{|\tilde{t}\rangle  | t = 0 ,1\}$ with $|\tilde{0}\rangle = |+\rangle$ and $|\tilde{1}\rangle = |-\rangle$. 
As the building block of MBQC, we note that for a pair of states entangled with $CZ$,
\begin{align}
{}_{1}\langle \widetilde{t} | 
Z^{s}_1 e^{i \theta Z_1} 
CZ_{1,2} |\psi\rangle_1 \otimes |+\rangle_2
= \frac{1}{\sqrt{2}} \Big( H e^{i \theta Z} Z^{t+s} \Big)_2 |\psi\rangle_2 . 
\end{align}
This involves a teleportation and a rotation gate.
The Pauli $Z^{t+s}$ depends on the measurement outcomes and is an example of by-product operators.
The set of unitary gates for MBQC is obtained by sequentially applying this formula to the rotated brickwork state $ |\psi_{bw}({\pmb \theta};{\pmb s}) \rangle$. 
It is convenient to define $w_v=t_v \oplus s_v \in \{0,1\}$ ($\oplus: \mathbb{Z}_2$ sum) for $v \in V$. 

\subsection{CNOT gate}

We look at the measurement pattern for the CNOT gate.
We measure qubits in $V$ except $5$ and $10$, where the output state will be defined.
(The expression $|\psi_{bw}(\pmb{\theta};\pmb{s})\rangle$ contains teleported phase gates acting on 5 and 10. 
They are used in subsequent steps in MBQC, where the output state induced at 5 and 10 is seen as a new input state. 
In the present analysis, we simply ignore them when we look at derived unitaries.)
We set
\begin{align}
&\theta_i = 0 \quad (i= 1,2,4,6,8)  , \nonumber \\ 
&\theta_9 = \alpha , \ \theta_3 = \beta , \ \theta_7=\gamma .
\end{align}
Using the formula, we get the following unitary acting on $|\psi\rangle_{5,10}$:
\begin{align}
&CZ \, (H Z^{w_4} \otimes H e^{ i\alpha Z } Z^{w_9} )
(H e^{i\beta Z } Z^{w_3} \otimes H  Z^{w_8} ) \nonumber \\ 
&\,CZ \,
(H Z^{w_2} \otimes H e^{i \gamma Z } Z^{w_7} )
(H  Z^{w_1} \otimes H  Z^{w_6} ) . 
\end{align}
Here, in $(\mathcal{P} \otimes \mathcal{Q})$, $\mathcal{P}$ acts on the qubit 5,  $\mathcal{Q}$ on the qubit 10.
It is equal to 
\begin{align}
\pm & \Big(X^{w_2 + w_4} Z^{w_1+w_3+w_9} \otimes X^{w_7 + w_9} Z^{w_4+w_6+w_8} \Big) \nonumber  \\
&\times \exp[ i (-1)^{w_2} \beta Z \otimes I ]
\exp[ i (-1)^{w_2+w_6+w_8} \alpha Z \otimes X ] \nonumber \\
& \times \exp[i (-1)^{w_6} \gamma I \otimes X ] .
\end{align}
The set of Pauli operators in front of rotation gates is an example of by-product operators in MBQC. 
It is always propagated to the front of unitaries we wish to simulate, and the set of parameters such as $(\alpha,\beta,\gamma)$ are chosen based on the preceding measurement outcomes.

To be more precise, if we regard the 10-qubits brickwork state as a block in the middle of MBQC, the ``initial" state $|\psi\rangle_{1,6}$ also carries by-product operators from former MBQC steps.
Thus we write the state defined at $1,6$ as $ |\psi\rangle_{1,6} =(X^{x_1} Z^{z_1} \otimes X^{x_6} Z^{z_6}) |\psi' \rangle_{1,6} $.  
Then the state after the set of measurement will be
\begin{align}
\pm & \Big(X^{w_2 + w_4 + x_1} Z^{w_1+w_3+w_9 + z_1} \nonumber \\ 
 & \quad \otimes X^{w_7 + w_9 + x_6} Z^{w_4+w_6+w_8+z_6} \Big) \nonumber  \\
&\times \exp[ i (-1)^{w_2+x_1} \beta Z \otimes I ] \nonumber \\
& \times \exp[ i (-1)^{w_2+w_6+w_8+ x_1 + z_6} \alpha Z \otimes X ] \nonumber \\
& \times \exp[ i (-1)^{w_6+z_6} \gamma I \otimes X ] 
|\psi' \rangle_{5,10} .
\end{align}
Now, let us choose the parameters as follows:
\begin{align} \label{eq:param-cnot}
\alpha &= (-1)^{w_2+w_6+w_8+ x_1 + z_6}  \times \frac{- \pi}{4} , \nonumber \\
\beta &= (-1)^{w_2+x_1}  \times \frac{ \pi}{4} ,  \nonumber \\
\gamma &= (-1)^{w_6+z_6} \times \frac{ \pi}{4} . 
\end{align}
Then we obtain the unitary
\begin{align}
\exp[ - \frac{\pi i}{4} (I-Z_5) (I-X_{10}) ] = CX_{5,10}
\end{align}
with the by-product operators and a constant phase in front.

We note that the parameters can be always set as in \eqref{eq:param-cnot} because all the measurement outcomes $w_i$ (as well as $x_i,z_i$) are obtained before the step to implement the parametric rotation.
For example, we have results of measurements $t_v, s_v$ (as well as $x_i,z_i$) at $\{1,2,3,6,7,8\}$ before the measurement with $\alpha$ at 9.
It is clear that in order for Alice to properly choose $\theta_i$ to be sent, she needs to know the preceding measurement outcomes by Bobs, which we denoted $t_v$, so she can construct $w_v = s_v \oplus t_v$ combined with her $s_v$.

\subsection{Single-qubit rotation gate}

With the measurement pattern with $\theta_4 = \theta_9 = 0 $, we find the following unitary:
\begin{align}
\pm
& \Big( X^{w_2+w_4 + x_1} Z^{w_1+w_3+w_9+z_1}  \nonumber \\
&\quad \otimes X^{w_7 + w_9 + x_6} Z^{w_4 + w_6+w_8 + z_6} \Big) \nonumber \\
\times & \Big(  \exp[i (-1)^{x_1} \theta_3 Z ] \exp[i (-1)^{w_1+z_1} \theta_2 X ] \nonumber \\
&\quad \times  \exp[i (-1)^{x_1} \theta_1 Z ]   \nonumber \\
& \quad \otimes  \exp[i (-1)^{x_6} \theta_8 Z ]  \exp[i (-1)^{w_6+z_6} \theta_7 X ] \nonumber \\
&  \quad \times \exp[i (-1)^{x_6} \theta_6 Z ]   \Big) .
\end{align}
Choosing the parameters as 
\begin{align}
 &\theta_1 = (-1)^{x_1} \alpha , \quad 
  \theta_2= (-1)^{w_1+z_1} \beta  , \quad
\theta_3 = (-1)^{x_1} \gamma  \\
 &\theta_6 =  (-1)^{x_6} \gamma' , \quad 
 \theta_7 =  (-1)^{w_6+z_6}  \beta' , \quad
\theta_8 = (-1)^{x_6} \alpha' , 
\end{align}
we obtain the single-qubit ration gate
\begin{equation}
R(\alpha,\beta,\gamma) = e^{i \gamma Z}  e^{i \beta X}  e^{i \alpha Z}   
\end{equation}
acting on 5 and 10 as $R(\alpha,\beta,\gamma) \otimes R(\alpha',\beta',\gamma')$.

\end{document}